\documentclass[journal,12pt,onecolumn, doublespace]{IEEEtran}
\ifCLASSINFOpdf
\else
\fi
\usepackage{graphicx}
\usepackage{cite}
\usepackage{array}
\usepackage{amsmath}
\usepackage{amssymb}
\usepackage{textpos}
\usepackage{subfigure}
\usepackage{enumerate}
\usepackage{enumitem}
\usepackage{lscape}
\usepackage{float}
\usepackage{pdfsync}
\usepackage{url}
\usepackage{color}
\usepackage{threeparttable}
\usepackage{multirow}
\usepackage{hyperref}
\usepackage{amsthm}
\usepackage{algorithm}
\usepackage{algorithmic}
\usepackage[latin1]{inputenc}
\usepackage{tikz}
\usetikzlibrary{shapes,arrows}

\newtheorem{theorem1}{Theorem}

\newtheorem{lem1}{Lemma}
\usepackage[font={footnotesize,singlespacing=1.0}]{caption}

\newcommand{\p}{\mathbb{P}}

\renewcommand{\d}{{\rm d}}

\newcommand{\bs}{{\rm BS}}

\newcommand{\mm}{{\rm mm}}

\renewcommand{\r}{{\rm R}}
\renewcommand{\d}{{\rm d}}

\newcommand{\LOS}{{\mathcal{L}}}
\newcommand{\NLOS}{{\mathcal{N}}}

\newcommand{\cachesizemm}{{\rm{C}}_m}
\newcommand{\cachesizemu}{\rm{C}_\mu}

\newcommand{\E}{\mathbb{E}}

\begin{document}
\title{An Analysis on Caching Placement for  Millimeter/Micro Wave Hybrid Networks}
%
\author{
\IEEEauthorblockN{Sudip Biswas, Tong Zhang, Keshav Singh,\\ Satyanarayana Vuppala, and Tharmalingam Ratnarajah}
\thanks{\hrulefill}
\thanks{Sudip Biswas, Tong Zhang, Keshav Singh, and Tharmalingam Ratnarajah are with Institute for Digital Communications, School of Engineering, University of Edinburgh, Edinburgh, UK. Email: \{sudip.biswas, k.singh, t.zhang, t.ratnarajah\}@ed.ac.uk.}
\thanks{Satyanarayana Vuppala is with University of Luxembourg, Luxembourg. Email: satyanarayana.vuppala@uni.lu.}
\thanks{Correspondent author: Tong Zhang. }
\vspace{-3em}}

\maketitle
\begin{abstract}
In this paper, we consider a hybrid millimeter wave (mmWave) and micro wave ($\mu$Wave) network from the perspective of \emph{wireless caching} and study the optimal probabilistic content/file caching placement at desirable base stations (BSs)
using a stochastic geometric framework.
 Considering the average success probability (ASP) of file delivery as the performance metric, we derive expressions for the association probability of the typical user to the mmWave and $\mu$Wave networks. Accordingly, we provide an upper bound for the ASP of file delivery and formulate the content caching placement scheme as an optimization problem with respect to caching probabilities, that jointly optimizes the ASP of file delivery considering both content placement and delivery phases. In particular, we consider the caching placement strategy under both noise-limited and interference-limited environments. We numerically evaluate the performance of the proposed caching schemes under essential factors, such as blockages in the mmWave network, cluster radius, BS density, and path loss and compare it with uniform caching placement, caching $M$ most popular contents, and random caching placement. Numerical results demonstrate the superiority of the proposed caching scheme over others, albeit certain trade-offs. 
\end{abstract}
	\vspace{-1.5em}
\begin{IEEEkeywords}\vspace{-1.2em} 
Wireless caching, millimeter-wave networks, micro wave networks, Poison point processes.\vspace{-1.0em}
	\end{IEEEkeywords}
\vspace{-.5em}
\section{Introduction}\vspace{-.5em}
\label{SEC_Intro}
Demand for higher capacity and lower latency in wireless networks is increasing exponentially, which has resulted in the development of the fifth generation (5G) wireless communication systems, with key goals of data rates in the range of Gbps, billions of connected devices, lower latency, improved coverage and reliability, and environment-friendly operation. A large portion of the data traffic is due to the surge in internet usage, mobile applications usage, social media and online video streaming through mobile devices (e.g. mobile phones, tablets, laptops, etc.). Accordingly in \cite{Erricson}, 
Ericsson has predicted that global mobile data traffic will surpass 100 ExaBytes by 2023.
%
Nonetheless, it is interesting to note that a substantial amount of the data traffic are redundantly generated over networks \cite{Golrezaei_2013}, as several popular contents are asynchronously {and} repeatedly requested by many users. Motivated by this, pre-fetching some popular video contents in the local caches of base stations (BSs), also termed as { wireless edge caching},
 has been considered as a promising technique to alleviate network traffic loads. In particular,  wireless edge caching has the advantages of $1)$ alleviating the burden of the backhaul by avoiding repeated transmission of the same contents from the core network to end-users, $2)$ reducing latency by shortening the communication distance,
{$3)$ improving network capacity and throughput, and $4)$ reducing the operational cost due to lower cost of storage memory than bandwidth.}

Further to alleviate the spectrum crunch and meet the data rate demands, while on one hand, techniques like heterogeneous networks (HetNets) and network densification through small cell networks (SCNs) \cite{bhushan2014network} have become commonplace in current wireless communications research,
%
on the other hand, millimeter wave (mmWave) frequencies are being considered as an alternative solution to the currently operational micro wave ($\mu$Wave) networks. Also, mmWave transmission by nature is more suitable for operation in small cells due to its limited range of propagation.

Hence, it is realistic to say that in future 5G networks, mmWave and traditional $\mu$Wave networks will exist in conjunction with each other, which will allow for a seamless coverage and transmission. Consequently, with regards to the aforementioned technologies, researchers are looking into how to maximize the average success probability (ASP) of content delivery in cellular networks to increase the users' quality of experience (QoE) \cite{song2015cachingplacement} \cite{chae2016caching}.
In this paper, we will consider a hybrid {mm/$\mu$Wave}
 network and demonstrate the feasibility of caching in such a network.
%


\subsubsection{Related work on wireless caching}

While caching has been used to maintain internet traffic over the last two decades, but mainly related to computer-based technologies \cite{kohonen2012content,ananthanarayanan2012pacman}, recently wireless caching has triggered considerable interest from both academia and industry due to its potential of reducing backhaul loads, latency and cost.
%
 In this regard, \cite{paschos2016wireless, bastug2014living} showed the effectiveness of proactive caching on the network edge to help reducing traffic congestion in backhaul links. 
While a similar system, but involving stochastic geometry based framework was studied in \cite{Mitici2013} for the scenario of BSs located in the Euclidian plane, caching in device-to-device (D2D) communications  was considered in \cite{altieri2015fundamental} and \cite{ji2016fundamental}. 
Further, the authors in \cite{Tamoor2016} proved that popularity-based optimal caching placement in terms of outage probability gives better results than uniform caching where all $\bs$s uniformly fetch all contents in their local caches irrespective of the popularity. 
{Hence, caching placement {scheme} is a key factor that determines the success and performance of a caching system, and it is an endeavour for researchers to find optimal ways to perform content caching placement in various networks. In this regard, while the authors in \cite{Shariatpanahi2014} explore the optimal methods for caching in a multiple-input multiple-output (MIMO) network, \cite{Golrezaei2014} studies optimal caching strategies for D2D communications.

\subsubsection{Related work on mmWave and hybrid networks}
Due to the vast amount of unused spectrum in mmWave frequency range, considerable efforts have been made to analyze mmWave systems recently. However, signals transmitted at mmWave frequencies are easily attenuated by blockages, such as concrete buildings, trees, etc.  Hence, to quantify the performance of mmWave cellular systems, \cite{MyListOfPapers:Akdeniz_GolbeCom} performed real time propagation channel measurements. Then, while in \cite{MyListOfPapers:Bai_TWC_2014} a blockage model for mmWave propagation was introduced to analyze the rate and coverage area of mmWave systems,  a distance dependent path loss model along with antenna gain parameters were introduced in \cite{MyListOfPapers:Singh_JSAC_2015} to characterize the propagation environment in such systems. 
Further, while the above works focussed on urban environments, \cite{maccartney2017rural} proposed a  path loss model for rural mmWave networks { and \cite{turgut2016coverage} showed the feasibility and application of mmWave in outdoor cellular networks.} 

On a similar vein, hybrid wireless networks, which combine two different tiers of base stations can also increase the capacity of wireless networks and spectrum efficiency. Accordingly, research works in \cite{1208989, 1402592, 1285914, 4769387, 7855640} have focussed on the study of capacity for such networks. 
Further, with regards to future 5G networks, the authors in \cite{7505974} were among the first to consider a mmWave overlaid $\mu$Wave network. Subsequently, the authors in \cite{elshaer2016downlink} provided uplink and downlink coverage analysis of such hybrid networks and \cite{turgut2016coverage} provided a coverage analysis in a {densified} 
heterogenous mmWave {SCN}.

Based on the aforesaid, in this paper we study the optimal content caching placement strategy in a mm/$\mu$Wave {hybrid }network
by maximizing the {ASP} of file delivery. In particular, considering a stochastic geometric framework, we model a hybrid cellular network involving both mmWave and $\mu$Wave BSs, which are randomly located within a 
bounded region. The BSs have finite cache memory and store
files in them in a probabilistic and independent way. We consider a typical user, which receives the content
of interest from the serving BSs offering the best received signal power among its neighbouring BSs.
The main distinctions of this work can be summarized as in the following points:
\begin{itemize}
\item We calculate the association probability of the user to a mmWave or $\mu$Wave network based on {the simplified long-term average biased (LTAB) received signal power (\emph{i.e.,} least path loss)}.
Using the association probability, we provide the expression for the ASP of file delivery. While, a closed-form expression is provided for the ASP of file delivery under a noise-limited scenario, an upper bound is provided for the interference-limited case.
\item To optimally place the contents in the hybrid network, we propose two algorithms, one each for noise-limited and interference-limited scenarios to acquire optimal caching probabilities by maximizing the ASP of file delivery. 
\item Besides characterizing the effects of blockages on the ASP for the typical user over a bounded region, we also evaluate the effects of channel fading in conjunction with other essential factors, such as mm/$\mu$Wave BS density, cluster radius, and content popularity. 
\item Finally, through numerical simulations we compare the performance of the proposed caching {placement }scheme with 
other caching strategies such as, $1)$ uniform caching placement, $2)$ caching $M$ most popular contents, and $3)$ random caching placement. Numerical results demonstrate the superiority of the proposed caching scheme over others, although with certain performance trade-offs. 
\end{itemize}
}
\vspace{-1.5em}
\section{System Model }\vspace{-0.5em}
		\begin{figure} [t!] 
			\centering
			\includegraphics[scale=0.5]{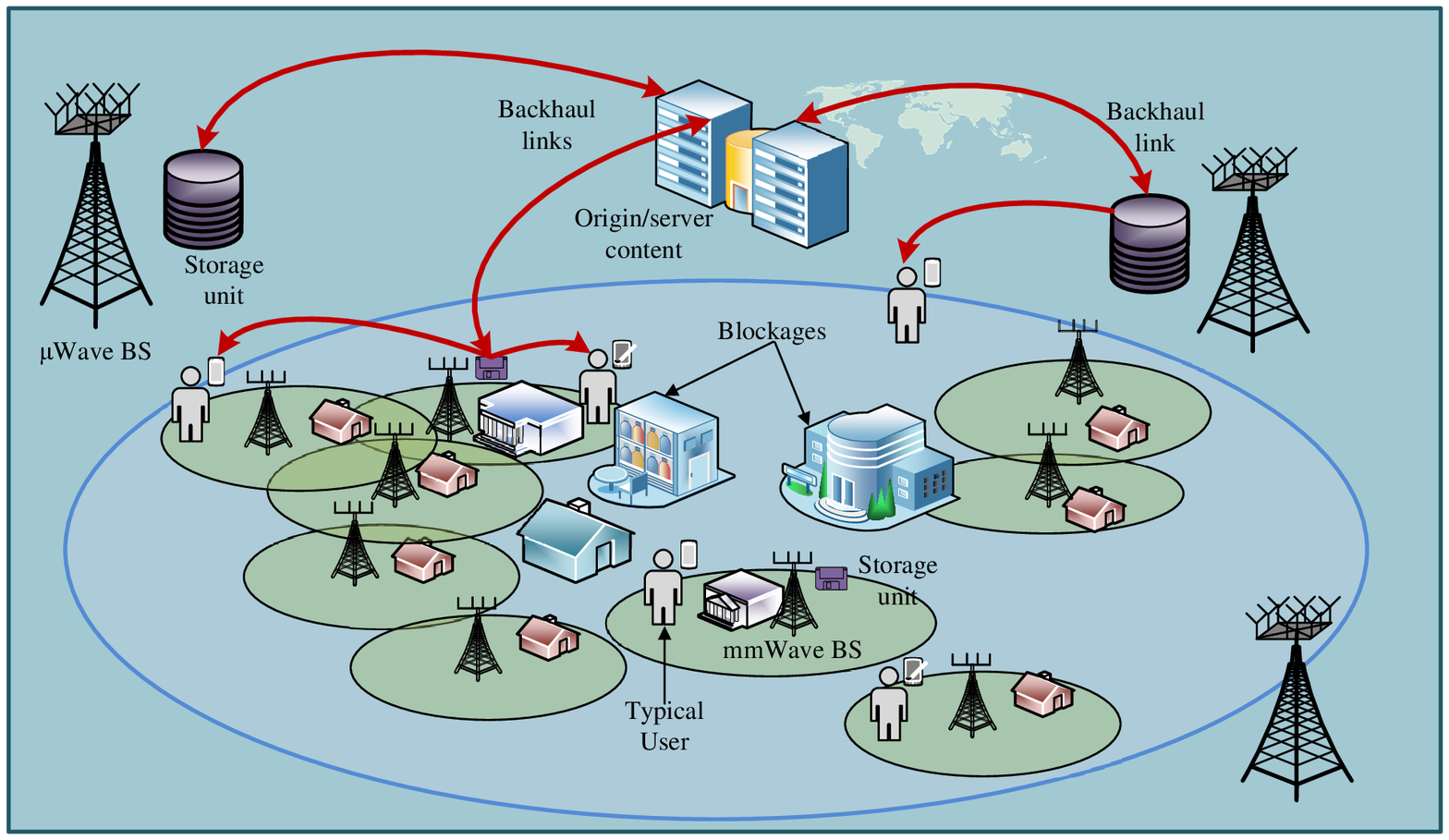}
			\caption{An illustration of a cache enabled hybrid mmWave-$\mu$Wave network.}
				\vspace{-3ex}
			\label{Sys_model}
		\end{figure}
We consider the downlink transmission in a cache-enabled hybrid cellular network comprising of both mmWave and $\mu$Wave networks as shown in Fig. \ref{Sys_model}. While  mmWave  BSs {and associated users }form the SCN{s},  $\mu$Wave BSs {and their associated users} form the macro cell network{s (MCNs)}.
Further,  mmWave BSs {and $\mu$Wave BSs} are {independently} modeled {by two
homogeneous Poisson point processes (PPPs) $\Phi_m$ with density $\lambda_m$ and
$\Phi_{\mu}$ with density $\lambda_{\mu}$, respectively}.  The users in the network follow another independent homogenous PPP $\Phi_u$ with density $\lambda_u$. All the processes are independent of each other. Further, both mmWave and $\mu$Wave BSs are equipped with multiple antennas $n_t^{m}$ and $n_t^{\mu}$, respectively. The users are assumed to be equipped with two sets of antennas $n_r^m$ and $n_r^{\mu}$, to receive both mmWave and $\mu$Wave transmissions, respectively\footnote{While to receive mmWave signals, the users utilize multiple antennas, $\mu$Wave signals are captured by a single antenna only. This is due to the fact that the wavelength of mmWave signals is very small and hence more antennas can be accommodated within a small physical space. However, for $\mu$Wave receivers, it may not be feasible to equip small devices with more than one antenna owing to larger wavelengths of $\mu$Wave signals. Henceforth, we use $n_r^{\mu}=1$.}. Hereinafter, throughout the paper subscript/superscript/notation of $m$ and $\mu$ will be used to refer to mmWave BSs and $\mu$Wave BSs{,} respectively.
%
In a typical cellular network, BSs retrieve requested files using capacity-limited backhaul links. During peak hours, this results in an information-congestion bottleneck both at the BSs as well as in the backhaul links. To alleviate this bottleneck, caching popular contents at the mmWave and $\mu$Wave BSs are proposed. The requested content will be served directly to the users by one of the neighbouring BSs depending on the availability of the file in its local cache and the association criteria of the users to the BSs. The performance of caching however, depends on the density of BS nodes, cache size, {users' request rate}, the caching strategy, content popularity, cluster radius and blockages in the network.
\vspace{-1.75em}
\subsection{Caching policy}\vspace{-0.75em}
Without loss of generality, a typical user is assumed to be located at the origin that is the centre of a {two dimensional (2-D) disk} with a radius $R$. This circle signifies a {finite network region},
 in which all BSs from both mmWave SCN and  $\mu$Wave MCN can communicate with { the typical user}.
%
The typical user will be associated to either mmWave  network or $\mu$Wave network depending on the {LTAB}
 received signal power and the availability of requested files at the serving BSs. In particular, the typical user is associated to the mmWave network if and only if 1) the strongest {LTAB}
 received signal power achieved at the associated mmWave BS is greater than the maximum value of the same at the $\mu$Wave BS, 2) the requested file is prefetched in the associated mmWave BS, selected according to the best received signal power, and 3) the rate supported by the serving mmWave BS is greater than the guaranteed target rate. If either of the aforementioned conditions cannot be satisfied by the mmWave network, the typical user will be offloaded to the $\mu$Wave network\footnote{Similar offloading strategies with respect to a threshold capacity were analyzed in \cite{MyListOfPapers:Singh_JSAC_2015, 6287527} and 
stated to be reasonable for mmWave based networks.}, where a potential $\mu$Wave serving BS will be chosen from a subset that has the requested file stored in its cache. Now, if the serving $\mu$Wave BS can achieve the guaranteed target rate, the requested file can be successfully delivered to the typical user. As for the case that the requested file is delivered from the local caches, it will be termed as ``cache hit". Otherwise, it is a ``cache miss", where BSs will then utilize backhaul links to retrieve the requested file in real time\footnote{The cache miss scenario is mentioned here for the sake of completeness, but is beyond the scope of discussion in this work.}. This scenario is not ideal for caching and should be avoided. Our aim is to develop an effective cache placement strategy to maximize ``cache hit". More details on the communication policy will be discussed in later sections.

\subsubsection{Caching model}
To cache popular files requested by a user, each BS contains storage units, referred to as local caches. The cache size of the $\mu$Wave  BSs is assumed to be larger than that of mmWave BSs. 
Additionaly, a central source/server containing a global cache\footnote{In the event that the file requested by the typical user is not in the local cache, then the  file is retrieved from global cache that contains all the files a user may request.} is accessible to all the BSs in the hybrid network via wired backhaul links. 
%
%
For simplicity, we represent the size of the cache by the number of files.
It is assumed that each mmWave and $\mu$Wave BS can cache up to $\cachesizemm$ and $\cachesizemu$ files
of length $B$ bits each, respectively, such that $\cachesizemm<\cachesizemu$. 
Further, 
{we assume that the distribution of users' requests follows the independent reference model (IRM), in which the content popularity is stationary and each user independently requests a data-file of size $B$ bits\footnote{For analytical simplicity, we assume that all files have the same size.} from the file set $\mathcal{S} \triangleq\{ s_1, s_2, \dots,s_i,\dots s_L \}$, where $L$ is the total number of files cached in the network. 
The popularity of the requested files {is} assumed to be independent of each other and is modelled by the Zipf distribution \cite{penrose2003random}.
In particular, the popularity of the $i$th file in the library is given as\vspace{-1.0em}
\begin{equation}
\label{eqn:zipf}
\begin{array}{lr}
f_i=\dfrac{{1}/{i^\upsilon}}{\sum_{j=1}^{\textit{L}} {1}/{j^\upsilon}},& 1\le i\le \textit{L}, 
\end{array}
\end{equation} 
where $\upsilon$ is the Zipf exponent, that controls the skewness of the content popularity. 
%
\subsubsection{Caching placement strategy}
A probabilistic caching placement {strategy} is assumed, where each BS (mm or $\mu$Wave) caches its file in an independently and identically distributed (i.i.d) manner by producing $M$ indices generated according to 
$\Pi \triangleq \{\pi_i:s_i\in \mathcal{S}, i = 1, 2, . . . , L\}$, where $ 0\leq\pi_i\leq1 \textrm{ and }\sum_{i=1}^L\pi_i\leq M$. 
%
The files are cached in advance {during off-peak hours} through prior requests or overhearing. The caching BSs storing file $i$ can be modelled as an independent PPP with intensity $\lambda_{j_i}\triangleq p_{j_i}\lambda_j$, where $p_{j_i}$ is the probability of caching the $i$th file for all $s_i\in \mathcal{S}$ and $j\in\{m,\mu\}$. Now, from {all} $L$ files, the typical user requests one file depending on the file popularity $f_i$, such that
a file with higher popularity is requested with higher likelihood. 
For analytical tractability, hereinafter we assume that the popularity of the files is perfectly known and that the files in the set $\mathcal S$ are of the same size and normalized to one\footnote{In the case of unequal file size, each file can be divided into small partitions of the same size, with each partition being treated as an individual file. }, \emph{i.e.,} $\mathcal{S} \triangleq\{1, 2, \dots, L \}$.
\vspace{-2.50em}
\subsection{MmWave network model}\vspace{-0.50em}
\subsubsection{Blockage model}
MmWave signals are susceptible to blockages, making it imperative to model blockages for true representation of practical mmWave systems. Blockages in the network are usually concrete buildings. We consider the blockages to be stationary blocks which are invariant with respect to direction. We adopt the modeling of blockages in \cite{MyListOfPapers:Bai_TWC_2015}, and accordingly, consider a two state statistical model for {each link}. The link can be either line-of-sight (LOS) or non-line-of-sight (NLOS).
LOS $({\LOS})$ link occurs when there is a direct propagation path between the transmitter and the receiver, while NLOS $({\NLOS})$ occurs when the link is blocked and the receiver receives the signal through reflection from a blockage. Let the LOS link be of length $r$ and $\beta$ be the blockage density, then the probabilities of occurrence $p_{\LOS}(.)$ and $p_{\NLOS}(.)$ of LOS and NLOS states, respectively, can be given as a function of $r$ as\vspace{-0.75em}
\begin{equation}
\label{EQ_ppp16b}
p_{\LOS}(r) =  e^ {-\beta r},\,\,\,\, p_{\NLOS} (r) =1- e^ {-\beta r}.\vspace{-1.0em}
\end{equation}
%
\subsubsection{Beamforming model} 
Due to the small wavelengths of mmWaves, directional beamforming at both transmitters and receivers can be exploited for compensating the path loss and additional noise. The beam patterns are approximated as sectorized gain patterns \cite{MyListOfPapers:Thornburg_TSP_2014}.
Let $\theta$ be the beamwidth of the main lobe. Then the antenna gain pattern for a transmit or receive node about an angle $\phi$ is given as \cite{MyListOfPapers:Thornburg_TSP_2014}

\begin{equation}
\label{EQ_ppp14}
G_q (\theta) = \left\{ 
\begin{array} {l l}
G_q^{\rm{M}} & \quad \mathrm{if |\phi| \leq \theta}\\ 
G_q^{\rm{m}} & \quad \mathrm{if |\phi| > \theta}
\end{array} \right\},
\end{equation}
where  $q \in  \{\rm{T},\r\}$, with $\rm{T}$ denoting the transmitter, and $\r$ the receiver, $\phi\in [0, 2\pi)$ is the angle of boresight direction and $G_q^{\rm{M}}$ and $G_q^{\rm{m}}$ are the array gains of main and side lobes, respectively. 
The effective antenna gain/interference\footnote{In order to avoid ambiguity of notation, hereinafter $G$ always denotes the effective antenna gain/interference.} seen by the typical user will depend on the directivity of the gains of main (\emph{i.e.,} $G^{\mathrm{M}} $) and side (\emph{i.e.,} $G^{\mathrm{m}} $) lobes of the antenna beam pattern, given as
	\begin{equation}
	\label{EQ_gains}
	G_i = \left\{ 
	\begin{array} {l l}
	G^{\mathrm{M}} G^{\mathrm{M}}, & \quad p_{\rm{MM}} = (\tfrac{\theta}{2 \pi})^2\\
	G^{\mathrm{M}} G^{\mathrm{m}}, & \quad p_{\rm{Mm}} = \tfrac{\theta (2 \pi-\theta)}{(2 \pi)^2}\\
	G^{\mathrm{m}} G^{\mathrm{M}}, & \quad p_{\rm{mM}} = \tfrac{\theta (2 \pi-\theta)}{(2 \pi)^2}\\
	G^{\mathrm{m}} G^{\mathrm{m}}, &  \quad p_{\rm{mm}} = (\tfrac{2 \pi -\theta}{2 \pi})^2
	\end{array} \right\},
	\end{equation}
where $p_{lk}$, with $l,k \in \{\mathrm{M,m}\}$, denotes the probability that the antenna gain $G^{l} G^{k}$ is seen by the typical user. Thus, the effective gain can be considered as a random variable, which can take any of the mentioned values in \eqref{EQ_gains}. For simplicity and tractability, we assume that the beamforming gain between the  mmWave serving BS and the typical user is always   $G^MG^M$ {since} the aligned beamforming gain between the typical user and the  mmWave serving  BS is maximum \cite{MyListOfPapers:Sudip_JSASP_2016}.
\subsubsection{Channel model}
%
%
To capture a generalized propagation environment and for analytical tractability, in this work we consider the Nakagami fading model\footnote{{The choice of Nakagami-$\hat{m}$ fading to simulate the small scale fading is commonly used in literature \cite{MyListOfPapers:Akdeniz_TJSC_2014,MyListOfPapers:Bai_TWC_2015,MyListOfPapers:Thornburg_TSP_2014}}.}. Let $\hat{m}$ be the Nakagami fading parameter and $\Gamma(\hat{m})$ the gamma function. Then,
the channel power is distributed as\vspace{-0.5em}
\begin{equation}
\mathcal{X}_{\hat{m}} \sim f_{\mathcal{X}_{\hat{m}}}(x;\hat{m}) \triangleq \frac{{\hat{m}}^{\hat{m}}x^{\hat{m}-1}e^{-\hat{m}x}}{\Gamma(\hat{m})}.\vspace{-1.0em}
\label{eqNakagamiChanPower}
\end{equation}
%
\subsection{$\mu$Wave network model}
The $\mu$Wave channels are modeled in a similar way as that of its mmWave counterparts with the only exceptions that the antennas are now omnidirectional with transmitted signal power $\mathrm{P}_{\mu}$ and path loss exponent $\alpha_{\mu}$. It is to be noted that the blockage effects are not considered for $\mu$Wave systems due to low penetration loss of $\mu$Wave signals. 

Under the consideration of separate encoding scheme at each BS, BS $l$ sends an information symbol $s_l$ through a
linear beamforming vector\footnote{A maximal ratio transmit (MRT) precoding scheme is considered, i.e., $\mathbf{v}_l=\frac{\mathbf{h}_{1,l}}{||\mathbf{h}_{1,l}||_2}$. } $\mathbf{v}_l = [v_l^1,\dots,{v_l}^{n^\mu_t}]^T$ with unit norm, \emph{i.e.,} $||\mathbf{v}_l||_2=1, i\in\Phi_{\mu}$. Therefore, by a slight abuse of notation, the received signal at the typical user from the $\mu$Wave BS $l$ can be given as
\vspace{-0.75em}
\begin{equation}
y=\sqrt{\mathrm{P}_{\mu}}\mathbf{h}_{1,l}^H\mathbf{v}_lr_l^{-\alpha_{\mu}/2}s_l+\sum\nolimits_{\substack{i\in\Phi{\color{red}}_{\mu},i\neq l}}\sqrt{\mathrm{P}_{\mu}}\mathbf{h}_{1,i}^H\mathbf{v}_ir_i^{-\alpha_{\mu}/2}s_i+n_1,\vspace{-0.5em}
\end{equation}
where {$\mathbf{h}_{1,i}=[h^1_{1,i},\dots,h^{n^\mu_t}_{1,i}]^T$
 is the downlink channel between the $\mu$Wave BS $i$ to the typical user\footnote{The subscript 1 in $\mathbf{h}_{1,l}$ corresponds to the typical user.} 
and $n_1$ denotes the additive Gaussian noise at the typical user.}

%
\vspace{-1.0em}
\section{Rate Characterization for the Typical User}\vspace{-0.50em}
The typical user may be connected to either a mmWave BS or a $\mu$Wave BS depending on the communication policy, which will be discussed in the next section. In this section we characterize the rate of the typical user when it is connected to either a mmWave or a $\mu$Wave BS. 
Let $\Phi_{m_i}$ be the set of mmWave BSs, which have file $i$ in their local caches.  Given that a typical user is served by a mmWave BS (from $\Phi_{m_i}$) that contains the requested file and with the strongest received signal power at the user is given as
\begin{equation}\label{max_signal}
\zeta_{m_{1,i}}=\underset{ k\in\Phi_{m_i}}{\max} \left\{ \dfrac{\mathrm{P}_m G_{1,k}\mathcal{X}_{1,k}}{r_{1,k}^{\alpha_{j}}}\right\},
\end{equation} 
where $\mathrm{P}_m$, $G_{1,k}$ and $\mathcal{X}_{1,k}$ are the transmit power, directional antenna gain, and channel {power} coefficient at the typical user from the {associated} mmWave BS, respectively. Further, $r_{1,k}$ is the distance between the typical user and the  serving mmWave BS and $\alpha_j$ is the path loss exponent with $j \in \{\LOS,\NLOS\}$.
%
Now, let $\Phi^c_{m_i}$ (or $\Phi_{\overline{m_i}}$) be the set of interfering BSs, \emph{i.e.,} interference from all other mmWave BSs {without} file $i$ in their cache memory. Then, $\Phi^c_{m_i}$ is written as\vspace{-1.0em}
\begin{equation}
\Phi^c_{m_i}=\Phi_{m}\backslash\Phi_{m_i}.\vspace{-1.0em}
\end{equation}
Therefore, the SINR at the typical user, receiving  file $i$ from the serving mmWave BS $l$, where $l\in\Phi_{m_i}$, can now be defined as\vspace{-1.0em}
\begin{eqnarray} \label{eq: received_sinr}
\gamma_{m_{1,{l_i}}}  &\triangleq& \dfrac{\mathrm{P}_{m} {G_{1,l}} \mathcal{X}_{{1,l}} {r^{-\alpha_j}_{1,l}}}{\Bigg\{\underbrace{\sigma^2_{m_{1,l}}}_{A}+\underbrace{\sum \nolimits_{t \in \Phi^c_{m_i}} \mathrm{P}_{m} {G_{1,t}} \mathcal{X}_{{1,t}} {r^{-\alpha_j}_{1,t}}}_{B}\Bigg\} }. 
%
\end{eqnarray}
In the denominator above, $A$ represents the noise power at the typical user, $B$ is the interference\footnote{We assume that each user is served by only one BS and BSs storing the same file cooperate among each other so that interference towards the typical user is mitigated. Round-robin scheduling is used to select users to be served in each time slot. Accordingly, the typical user suffers from interference in the network only from the BSs that do not cache the requested file.}
seen at the typical user from the mmWave BSs which do not contain the file $i$.
%
Accordingly, the downlink rate\footnote{Hereinafter, for notational simplicity, we omit the subscript 1 that is used to represent the typical user. } at the typical user requesting the $i$th file is expressed as\vspace{-1.0em}
\begin{equation} \label{eq: downlink_rate}
\mathcal{R}_{m_{l_i}}=({1}/{N_{l_i}})\times\log_2 \left(1+\gamma_{m_{{l_i}}}\right), \vspace{-0.50em}
\end{equation} 
where $N_{l_i}$ is the total load of the  serving mmWave BS. Similarly, {by a slight abuse of notation, we use the same symbols to denote the total load of the associated $\mu$Wave BS and the distance between the typical user and its associated $\mu$Wave BS}. Accordingly, the downlink rate at the typical user, receiving file $i$ from the serving $\mu$Wave BS $l$, with $l \in \Phi_{\mu_i}$, is given as \vspace{-0.750em}
\begin{equation} \label{eq: downlink_muwaverate}
\mathcal{R}_{\mu_{l_i}}=({1}/{N_{l_i}})\times\log_2 \left(1+\gamma_{\mu_{{l_i}}}\right), \vspace{-1.0em}
\end{equation} 
where\vspace{-1.20em}
\begin{eqnarray}
\label{eq: received_sinr_mu}
\gamma_{\mu_{l_i}}  &\triangleq& \dfrac{\mathrm{P}_{\mu} {H_l} r^{-\alpha_\mu}_l}{\Big\{\underbrace{\sigma^2_{\mu_l}}_{A}+\underbrace{\sum\nolimits_{t \in \Phi^c_{\mu_i}} \mathrm{P}_{\mu} {H_t} r^{-\alpha_\mu}_t}_{B}\Big\}}.\vspace{-0.75em}
\end{eqnarray}
Here, $H_{l}=||\mathbf{h}^H_{l} \mathbf{v}_{l}||^2=||\mathbf{h}_l||^2$ denotes the power gain of the fading channel\footnote{
Ignoring the independence among $n^\mu_t$ channels and sacrificing the channel diversity gain, while assuming that each entity of $\mathbf{h}_l$ is an i.i.d. complex Gaussian random variable with zero mean and variance ${1}/{n^\mu_t}$, $H_l$ can be considered to be Nakagami-$\hat{m}$ distributed with $\hat{m}=1$. Henceforth, this simplification helps us to obtain tractable mathematical derivations.}
 Like before, $A$ and $B$ denote noise and interference from the set of $\mu$Wave BSs not having file $i$, respectively.
%
\vspace{-1.0em}
\section{Association Probability}\vspace{-0.5em}
In this section, we present the communication policy, where we determine the association probability of the typical user $p_{mw}$ and $p_{\mu w}$ connected to either a mmWave network or a $\mu$Wave network, respectively.
To calculate $p_{mw}$ and $p_{\mu w}$, we formulate a problem by comparing the maximum {LTAB} received signal powers between mmWave and $\mu$Wave BSs. Accordingly, we introduce bias factors $B_\mu$ and $B_m$ \cite{6287527,6497002}, which are always positive. When $B=1$, no biasing is considered and the association goes back to a traditional cell association policy based on maximum received power or nearest node.  
%
Leveraging the analysis from \cite{6287527}, we consider that the typical user is connected to the best network with respect to {LTAB} received power (\emph{i.e.,} $B_m\mathrm{P}_mG_lr^{-\alpha_j}_l, j \in \{m,\mu\}$ for mmWave network and $B_\mu \mathrm{P}_\mu r^{-\alpha_\mu}_l$ for $\mu$Wave network). However, unlike $\mu$Wave network, it is important to characterize the least path loss distribution in  mmWave network by incorporating the effect of blockages. As mentioned before, the possibility of the channel link being LOS or NLOS follows from the exponential blockage probability model. Accordingly, the least path loss distribution for a typical user in a mmWave network  is given in Lemma \ref{cdf}, followed by the association probability of the typical user to $\mu$Wave network in Theorem \ref{association}. A pictorial representation of the above  communication policy is presented in Fig. \ref{flow_chart}.
\begin{lem1}\label{cdf}
The least path loss distribution for a typical user in a mmWave network is given as
\begin{eqnarray}
\label{EQ_leastPathlossmmWave}
F^{\mm}_{\xi_l}(r)\!=& \!1\!-\! \exp\left(-\pi \lambda_m (r \mathrm{P}_m G_l B_m)^{\tfrac{2}{\alpha_\NLOS}} - \tfrac{2 \pi \lambda_m}{\beta^2}\! (1\!- \!e^{-\beta (r \mathrm{P}_m G_l B_m)^{\tfrac{1}{\alpha_\LOS}} } \!(1+\beta (r \mathrm{P}_m G_l B_m)^{\tfrac{1}{\alpha_\LOS}} )) \right. \nonumber \\
&  \left.\!\! \hspace{8em}+ \tfrac{2 \pi \lambda_m}{\beta^2}\! (1\!- \!e^{-\beta (r \mathrm{P}_m G_l B_m)^{\tfrac{1}{\alpha_\NLOS}} } \!(1+\beta (r \mathrm{P}_m G_l B_m)^{\tfrac{1}{\alpha_\NLOS}} )) \right).
\end{eqnarray}
\end{lem1}\vspace{-0.75em}
\begin{proof}
The proof of this lemma follows from the proof of Theorem 1 of \cite{7370940}. However, for convenience, we present a sketch of the proof here. Consider a point process, where the points represent the path loss between the typical user and randomly placed BSs in a mmWave network. Let $\phi_\mm =\left\{\xi_l \triangleq \tfrac{x^{\alpha_m}_l}{\mathrm{P}_m G_l B_m}\right\}$ be a  homogeneous PPP of intensity $\lambda_m$. 
Here, the distance  is a random variable, and  its LOS state occurs with the probability of $e^{-\beta x}$.
By using Mapping theorem \cite[Theorem 2.34]{MyListOfPapers:HaenggiBook2012}, the density function of this one dimensional PPP  under the effect of blockages can be given as
\vspace{-1.0em}
\begin{eqnarray}
\label{EQ_totalPointsLOS}
\Lambda([0,r])= \int \limits_{0}^{(r \mathrm{P}_m G_l B_m)^{\tfrac{1}{\alpha_\LOS}}} 2 \pi \lambda_m x e^{-\beta x} \d x + \int \limits_{0}^{(r \mathrm{P}_m G_l B_m)^{\tfrac{1}{\alpha_\NLOS}}} 2 \pi \lambda_m x (1-e^{-\beta x} )\d x.
\end{eqnarray}

Using the void probability of a PPP and with the help of \eqref{EQ_totalPointsLOS},  the least path loss distribution  in  a mmWave network can be given as \eqref{EQ_leastPathlossmmWave}, where $B_m=1/\mathrm{P}_m$ and $B_\mu=1/\mathrm{P}_\mu$.

\tikzstyle{decision} = [diamond, draw, fill=white!20, text width=6em, text badly centered, node distance=3cm, inner sep=0pt]
\tikzstyle{block} = [rectangle, draw, fill=white!20, text width=9em, text centered, rounded corners, minimum height=4em]
\tikzstyle{line} = [draw, -latex']
\begin{centering}
\begin{figure}[t!]
  \tiny
 \hspace*{14.25em}
  \begin{tikzpicture}[node distance = 2cm, auto]
    \node [block] (init) { User requests $i$th file};
    \node [decision, below of = init,node distance = 2cm] (pathloss) { mmWave or $\mu$-Wave network?};
    \node [block, left of = pathloss, node distance = 4cm] (associated-mmwaveBS) {Select the associated mmWave BS};
    \node [decision, below of = associated-mmwaveBS,node distance = 1.98cm] (file-mm) {Is requested file available?};
    \node [decision, below of = file-mm, node distance = 2cm] (rate-mm) { $\mathcal{R}_{m_i}\geq\nu_i$?};
    \node [block, below of = rate-mm, node distance = 1.65cm] (mm) {Successfully served by the mmWave BS};
    \node [block, right of = pathloss, node distance = 4cm] (associated-muwaveBS) { Select the associated $\mu$Wave BS};
    \node [decision, below of = associated-muwaveBS, node distance = 1.98cm] (file-mu) {Is requested file available?};
    \node [decision, below of = file-mu, node distance = 2cm] (rate-mu) {$\mathcal{R}_{\mu_i} \geq \nu_i$?};
    \node [block, below of = rate-mu, node distance = 1.65cm] (mu) { Successfully served by the $\mu$Wave BS};
    \node [block, below of = pathloss, node distance = 2cm] (backhaul) { Retrieve through backhaul link};
    \node [block, below of = backhaul, node distance = 2cm] (fail){ Failure in serving the user};
    \path [line] (init) -- (pathloss);
    \path [line] (pathloss) -- node{($p_{mw}$)}(associated-mmwaveBS);
    \path [line] (pathloss) -- node{($p_{\mu w}$)}(associated-muwaveBS);
    \path [line] (associated-mmwaveBS) -- (file-mm);
    \path [line] (file-mm) -- node {(Y)} (rate-mm);
    \path [line] (file-mm) -- node {(N)}(backhaul);
    \path [line] (rate-mm) -- node {(Y)} (mm);
    \path [line] (rate-mm) -- node {(N)}(fail);
    \path [line] (associated-muwaveBS) -- (file-mu);
    \path [line] (file-mu) -- node{(Y)}(rate-mu);
    \path [line] (rate-mu) -- node{(Y)}(mu);
    \path [line] (rate-mu) -- node{(N)}(fail);
    \path [line] (file-mu) -- node {(N)}(backhaul);
  \end{tikzpicture}
  \caption{An illustration of the communication policy for the typical user with either mmWave or $\mu$Wave network.}\vspace{-3.5em}
  \label{flow_chart}
\end{figure}
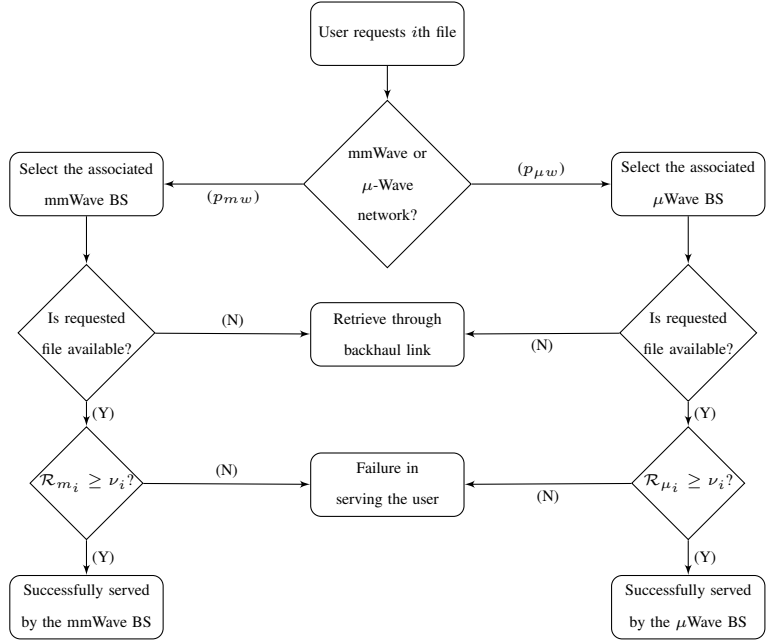
\end{centering}
\end{proof}
\begin{theorem1}
\label{association}
In a hybrid network consisting of mmWave and $\mu$Wave BSs as described, considering the least pass loss distribution, the probability that a  typical user is connected to the $\mu$Wave network is given by
\vspace{-1.0em}
\begin{equation}
\label{EQ_PDFcontactDistmmWave2}
p_{\mu w}  \!=\!  2\pi \lambda_\mu\! \!\int \limits_{0}^{\infty}\!\!  r \exp\left(\!-\Lambda_m\left(\left(\tfrac{\bar{\mathrm{P}}_\mm}{\bar{\mathrm{P}}_\mu } \!\right)^{\tfrac{1}{\alpha_m}} r^{\tfrac{\alpha_\mu}{\alpha_m}}\right)\!\right) \! \! e^{\!- \pi \lambda_\mu r^2}  \d r,
\end{equation}
where $\bar{\mathrm{P}}_\mm = \mathrm{P}_m G_l B_m$, with $G_l  = G^MG^M$ and $\bar{\mathrm{P}}_\mu = \mathrm{P}_\mu B_\mu$, with $B_m = \frac{1}{\mathrm{P}_m}$, $B_\mu = \frac{1}{\mathrm{P}_{\mu}}$. Then,
\vspace{-0.5em}
\begin{align}
\Lambda_m\left({\tfrac{\bar{\mathrm{P}}_\mm}{\bar{\mathrm{P}}_\mu } \!}^{\tfrac{1}{\alpha_m}} r^{\tfrac{\alpha_\mu}{\alpha_m}}\right) =&\pi \lambda_m \left(\tfrac{\bar{\mathrm{P}}_\mm}{\bar{\mathrm{P}}_\mu } \!\right)^{\tfrac{1}{\alpha_\NLOS}} r^{\tfrac{\alpha_\mu}{\alpha_\NLOS}}-\tfrac{2 \pi \lambda_m}{\beta^2} \!\!\left(\!\!1\!- \!e^{-\beta\left(\tfrac{\bar{\mathrm{P}}_\mm}{\bar{\mathrm{P}}_\mu } \!\right)^{\tfrac{1}{\alpha_\NLOS}} r^{\tfrac{\alpha_\mu}{\alpha_\NLOS}}}\!\!\!\left(\!1+\beta\left(\tfrac{\bar{\mathrm{P}}_\mm}{\bar{\mathrm{P}}_\mu } \!\right)^{\tfrac{1}{\alpha_\NLOS}} r^{\tfrac{\alpha_\mu}{\alpha_\NLOS}}\right)\right) \nonumber \\
&+ \tfrac{2 \pi \lambda_m}{\beta^2} \!\!\left(\!\!1\!- \!e^{-\beta\left(\tfrac{\bar{\mathrm{P}}_\mm}{\bar{\mathrm{P}}_\mu } \!\right)^{\tfrac{1}{\alpha_\LOS}} r^{\tfrac{\alpha_\mu}{\alpha_\LOS}}}\!\!\! \left(\!1+\beta\left(\tfrac{\bar{\mathrm{P}}_\mm}{\bar{\mathrm{P}}_\mu } \!\right)^{\tfrac{1}{\alpha_\LOS}} r^{\tfrac{\alpha_\mu}{\alpha_\LOS}}\right)\right). 
\end{align}
%
\end{theorem1}
\begin{proof}
The proof can be obtained by leveraging results of Lemma \ref{cdf}.
%
\end{proof}
%
\section{Performance metric}
We use the average success probability (ASP) of file delivery as the performance metric to design the caching placement strategy.
%
In particular, ASP is defined as the successful response to the user's request, which occurs when the downlink rate is more than the target bit rate of the file. Thus, when the typical user requests the $i$th file, the ASP of file delivery can be expressed as\vspace{-1.0em}
\begin{equation}
\mathcal{P}_s(\{\nu_i\})=\sum\nolimits_{i=1}^{\textit{L}} f_i \,\, \p\left[\mathcal{R}_{x_i} \ge \nu_i\right], \label{eq: success_probability}
\end{equation}
 where $x_i \in \Phi_{m_i/\mu_i}$ is the serving BS\footnote{For notational simplicity, $x_i$ denotes either $\mu$Wave serving BS or mmWave serving BS for file $i$ depending on the context.} selected according to \eqref{max_signal}, $f_i$ is the probability of requesting the $i$th file, $\nu_i$ is the target bit rate of file $i$ and $\mathcal{R}_{x_i}$ is the rate at the typical user.

Now, as the mmWave and $\mu$Wave networks follow two independent PPPs, it is possible to perform the analysis on both the processes independently with an {association probability}. 
Accordingly, {given that} $p_{mw}$ is the association probability that the typical user is connected to the mmWave network, the file $i$ {is served by} the associated mmWave BS that is able to support the downlink rate greater than the target bit rate. {Otherwise, the typical user will be associated to the $\mu$Wave network with the probability $p_{\mu w}=1-p_{mw}$}. In this regard, we assume that the typical user can communicate with all BSs that cache the requested file\footnote{If the user cannot be served from the local caches due to low delivery rate, other BSs meeting the target rate requirement, but without file $i$ in their local caches should retrieve the requested file via backhaul links or coordinated transmission between BSs. This however is out of scope of the current work and is left for future work.}. Accordingly, the total ASP of file delivery can be given as \vspace{-1.0em}
%
\begin{equation}
\mathcal{P}_s(\{\nu_i\})=\mathcal{P}^{\mm}_s(\{\nu_i\}) p_{mw}+\mathcal{P}^{\mu}_s(\{\nu_i\})  p_{\mu w},\vspace{-1.0em}
\end{equation}
where $\mathcal{P}^{\rm mm}_{s} (\nu_i) $ and $\mathcal{P}^{\mu}_{s} (\nu_i) $ denote the conditional ASP of file delivery by the mmWave and $\mu$Wave networks, respectively.\vspace{-0.75em}

\begin{theorem1}
\label{theorem mmwave}
The ASP of file delivery by the mmWave BSs is tightly upper bounded by\vspace{-0.250em}
\begin{align}
\mathcal{P}^{mm}_s(\{\nu_i\}) \leq&\sum\nolimits_{i=1}^{\textit{L}} \left\{f_i \left\{\sum\nolimits_{l=1}^{\hat{m}} \binom{\hat{m}}{l} (-1)^{l+1} \sum\nolimits_{j\in\{L,N\}} \left\{p_j \textup{exp}\Big(\dfrac{-AlQ_i\sigma^2_{m_{x_i}}}{\mathrm{P}_m G_{x_i} r^{-\alpha_j}_{x_i}}\Big)\right. \right.\right. \\ \times
&\left.\left.\left.\left\{\prod_{\hat{q}=1}^{3}\prod_{\hat{j} \in \{\LOS,\NLOS\}}  \textup{exp}\Big[-\overline{\lambda}_{m_i} \int_{0}^{\infty}\Big(1- \Big(\frac{1}{1+\frac{A l Q_i  \hat{G}_{\hat{q}}  r^{-\alpha_{\hat{j}}}_t}{G_{x_i} r^{-\alpha_{j}}_{x_i}\hat{m}}}\Big)^{\hat{m}}\Big) p_{\hat{j}} 2 \pi p_{\hat{q}} \, r_t  dr_t\Big]\right\}\right\}\right\}\right\}\nonumber\,,
\end{align}
where $G_{x_i} = G^MG^M, \hat{G}_{\hat{q}} \in \{\hat{G}_1 = G^MG^M, \hat{G}_2=G^{\rm{M}}G^{\rm{m}}, \Hat{G}_3 = G^{\rm{m}}G^{\rm{m}} \}$, $p_{\hat{q}} \in \{p_1 = p_{\rm{MM}}, p_2=2p_{\rm{Mm}}, p_3=p_{\rm{mm}}\}$ with $\hat{q} \in \{1,2,3\}$.
Particularly, $p_{j}, p_{\hat{j}} \in \{p_\NLOS=1-e^{-\beta r_{x_i}}, p_\LOS=e^{-\beta r_{x_i}}\}$ with $j,\hat{j} \in \{\LOS,\NLOS\}$, $A = \hat{m}(\hat{m}!)^{\frac{-1}{\hat{m}}}$, where for mmWave network, $\hat{m}>1$ is the Nakagami parameter. $Q_i=2^{N_{x_i}\nu_i}-1$, and $\overline{\lambda}_{m_i}=(1-p_{m_i})\,\lambda_m$.
\end{theorem1}
\begin{proof}
This proof is given in Appendix A.
\end{proof}
\begin{theorem1}\label{theorem_mu}
The ASP of file delivery by the $\mu$Wave BSs is tightly upper bounded by
\begin{align}
\mathcal{P}^{\mu}_s(\{\nu_i\})\leq&\sum\nolimits_{i=1}^{\textit{L}}f_i\left\{\sum\nolimits_{l=1}^{\hat{m}}\left\{\binom{\hat{m}}{l}(-1)^{l+1} \textup{exp} (\frac{-lAQ_i \sigma^2_{\mu_{x_i}}}{\mathrm{P}_\mu r^{-\alpha_\mu}_{x_i}})\right.\right.\nonumber \\ 
& \left.\left.\textup{exp}[-2\pi \overline{\lambda}_{\mu_i}\int_{0}^{\infty} r_t(1-(1+\frac{A l Q_i  r^{-\alpha_\mu}_t}{r^{-\alpha_\mu}_{x_i}\hat{m}})^{-\hat{m}})dr_t]\right\}\right\}\,,\vspace{-1.0em}
\end{align}
where $\hat{m} = 1$, $Q_i$\footnote{$N_{x_i}$ denotes the total load of either the associated mmWave or $\mu$Wave BS depending on the context. $Q_i$ for mmWave and $\mu$Wave are usually different as the total load of the serving BSs for file $i$ may be different. However, for simplicity we assume that the total loads are equal in this work.
}$ = 2^{N_{x_i}\nu_i}-1$ and $\overline{\lambda}_{\mu_i}=\lambda_\mu (1-p_{\mu_i})$. All parameters are either defined before or are generally the same as those defined for the mmWave network but with notational changes.
\end{theorem1}
\begin{proof}
Due to page limitations, the proof of this theorem is omitted but can be obtained in a similar way as the proof of Theorem \ref{theorem mmwave}.
\end{proof}
\vspace{-1.0em}
\section{Proposed Caching Placement in the Hybrid Network}\vspace{-0.5em}
To place the contents in the hybrid network, we aim to optimize the ASP of file delivery by considering a finite memory size and content popularity. This can be achieved by optimally determining caching probabilities in the file caching placement phase. Accordingly, we formulate an optimization problem as below.
\vspace{-1.0em}
\begin{align}
\,\,\textbf{P1}\,: \max_{\{p_{m_i}\},\{p_{\mu_i}\}}&\mathcal{P}_s(\{\nu_i\}),\qquad\label{eq: objective function1}\\ 
\text{s.t.}\,\,\, &\sum\nolimits_{i=1}^{L} p_{m_i}\leq \cachesizemm,\label{eq: lagrangian constraint1}\\
& \sum\nolimits_{i=1}^{L}  p_{\mu_i}\leq \cachesizemu,\label{eq: lagrangian constraint3}\\
& 0\leq p_{m_i}\leq1\;\;\text{\&}\;\;0\leq p_{\mu_i}\leq1 \,,\;\, \forall i \in \mathcal{S},\label{eq: lagrangian constraint2}
\end{align}
The constraints in~\eqref{eq: lagrangian constraint1} and \eqref{eq: lagrangian constraint3} ensure that the size of the total cached files should be less than or equal to $\cachesizemm$ for mmWave and $\cachesizemu$ for $\mu$Wave networks.  
%
In the above, the ASP is given as in~\eqref{eq58}, shown on the top of the this page.
\begin{figure*}
\begin{align}
&\mathcal{P}_s(\{\nu_i\}) = \mathcal{P}^{\mm}_s(\{\nu_i\}) p_{mw}+\mathcal{P}^{\mu}_s(\{\nu_i\}) p_{\mu w}\nonumber\\
&\leq\sum\nolimits_{i=1}^{\textit{L}} \left\{f_i \left\{\sum\nolimits_{l=1}^{\hat{m}} \binom{\hat{m}}{l} (-1)^{l+1} \sum\nolimits_{j\in\{\LOS,\NLOS\}} \left\{p_j \textup{exp}\Big(\dfrac{-AlQ_i\sigma^2_{m_{x_i}}}{\mathrm{P}_m G_{x_i} r^{-\alpha_j}_{x_i}}\Big)\right. \right.\right. \nonumber\\ 
&\times\left.\left.\left.\left\{\textup{exp}\Big[-\overline{\lambda}_{m_i} \underbrace{\sum\nolimits_{\hat{q}=1}^{3}\sum\nolimits_{\hat{j} \in \{\LOS,\NLOS\}}\int_{0}^{\infty}\Big(1- \Big(\frac{1}{1+\frac{A l Q_i  \hat{G}_{\hat{q}}  r^{-\alpha_{\hat{j}}}_t}{G_{x_i} r^{-\alpha_{j}}_{x_i}m}}\Big)^{\hat{m}}\Big) p_{\hat{j}} 2 \pi p_{\hat{q}} \, r_t  dr_t\Big]}_{Z(i,l)}\right\}\right\}\right\}\right\} p_{m w}\nonumber \\
&+\sum\nolimits_{i=1}^{\textit{L}}f_i\left\{\sum\nolimits_{l=1}^{\hat{m}}\left\{\binom{\hat{m}}{l}(-1)^{l+1} \textup{exp} (\frac{-lAQ_i \sigma^2_{\mu_{x_i}}}{\mathrm{P}_\mu r^{-\alpha_\mu}_{x_i}})\right.\right.\nonumber \\ 
& \left.\left.\textup{exp}[- \overline{\lambda}_{\mu_i}\underbrace{\int_{0}^{\infty} 2\pi r_t(1-(1+\frac{A l Q_i  r^{-\alpha_\mu}_t}{r^{-\alpha_\mu}_{x_i}\hat{m}})^{-\hat{m}})dr_t]}_{W(i,l)}\right\}\right\} p_{\mu w}\nonumber \\
&=\sum\nolimits_{i=1}^{\textit{L}} \left\{f_i \left\{\sum\nolimits_{l=1}^{\hat{m}} \binom{\hat{m}}{l} (-1)^{l+1} \hspace{-1em}\sum_{j\in\{\LOS,\NLOS\}} \left\{p_j \textup{exp}\Big(\dfrac{-AlQ_i\sigma^2_{m_{x_i}}}{\mathrm{P}_m G_{x_i} r^{-\alpha_j}_{x_i}}\Big) 
 \textup{exp}\Big[-\overline{\lambda}_{m_i} Z(i,l)\Big]\right\}\right\}\right\} p_{m w}\nonumber \\
&+\sum\nolimits_{i=1}^{\textit{L}}f_i\left\{\sum\nolimits_{l=1}^{\hat{m}}\left\{\binom{\hat{m}}{l}(-1)^{l+1} \textup{exp} (\frac{-lAQ_i \sigma^2_{\mu_{x_i}}}{\mathrm{P}_\mu r^{-\alpha_\mu}_{x_i}})\textup{exp}[- \overline{\lambda}_{\mu_i}W(i,l)]\right\}\right\} p_{\mu w}.\label{eq58}
\end{align}
\hrule
\vspace{-1.5em}
\end{figure*}
The {functions consisting of} variables $p_{m_i}$ and $p_{\mu_i}$ are exponential functions, which are convex. Hence, their summation is also convex. However, due to the binomial term $(-1)^{l+1}$, the objective function is no longer convex. In particular, it is a difference of convex (DC) functions and hence a {\it DC } (DCP), which makes it quite rigorous to cope with the optimization problem. In this regard, we consider the noise-limited (NL) and interference-limited (IL) scenarios separately to try to simplify the problem and reformulate the optimization problem under two different scenarios. 
\subsection{Optimal caching probabilities under NL scenario}
Recent studies on mmWave networks \cite{MyListOfPapers:Sudip_JSASP_2016}, state that mmWave networks in urban settings are more NL than IL. This is due to the fact that in the presence of blockages, the signals received from unintentional sources are close to negligible. In such densely blocked scenarios (typical for urban settings), signal to noise ratio (SNR) provides a good enough approximation to  SINR for directional mmWave networks. Furthermore, for the $\mu$Wave case, when the number of users is much smaller than the number of serving BSs, the impact of interference is negligible when compared to the noise power. Accordingly, the typical user can be served without sharing resources with other users. Hence, in this section we ignore the interference part $B$  in \eqref{eq: received_sinr} and \eqref{eq: received_sinr_mu}, and consider the NL case only to analyze the effects of channel selection diversity on optimal caching strategy. 
Additionally, such an assumption also aids us in deriving closed form expressions for the ASP of file delivery.
\begin{theorem1}\label{theorem mmwave1}
The ASP of file delivery by the mmWave network in a NL scenario is given by
\begin{align}
\mathcal{P}^{\textrm{mm}}_s(\{\nu_i\}) = \sum\nolimits_{i=1}^{L}f_i\Bigg[1 - \textup{exp}\bigg(-\sum\nolimits_{j\in\{\LOS,\NLOS\}}k_j p_{m_i} Z_j\big(\frac{\eta_i G_{x_i}}{Q_i }\big)-\hat{k}p_{m_i}\big(\frac{\eta_i G_{x_i}}{Q_i}\big)^{\delta_{m_\NLOS}}\bigg)\Bigg]\label{mmwave}
\end{align}
\end{theorem1}
where $k_j=c_j\frac{\pi\lambda_m {\hat{m}}^{\hat{m}}\delta_{m_j}}{\Gamma(\hat{m})}$, $c_j\in\{-1,1\}\footnote{$c_\LOS = 1, c_\NLOS = -1$.}
$, with $j \in \{\LOS,\NLOS\}$, $\hat{k}=\pi \lambda_m \frac{\Gamma(\delta_{m_\NLOS}+\hat{m})}{m^{\delta_{m_\NLOS}}\Gamma(m)}$, $\delta_{m_j}=\frac{2}{\alpha_j}$, $\eta_i=\frac{\mathrm{P_m}}{\sigma^2_{m_{x_i}}}$, $Q_i=2^{\rho_i}-1$, $\rho_i=N_{x_i} \nu_i$. $Z_j(\tilde{\omega})= \int_{0}^{\infty}\int_{0}^{\tilde{\omega}}\frac{\textrm{exp}(-\frac{\hat{m}}{\omega}\psi)}{\omega^{(\hat{m}+1)}}\textrm{d}\omega \psi^{(\delta_{m_j}+\hat{m}-1)} \textrm{exp}(-\beta \psi^{\frac{\delta_{m_j}}{2}})\textrm{d}\psi$ with $\tilde{\omega}=\frac{\eta_iG_{x_i}}{Q_i}$. All other parameters are as defined before.
\begin{proof}
The proof is given in Appendix C.
\end{proof}

Similarly, the ASP of file delivery in the $\mu$Wave network can be derived with the channel power gain now being {exponentially distributed}, {where}  blockage effects and beamforming gains are ignored.
\begin{theorem1}\label{theorem muwave1}
The ASP of file delivery in a $\mu$Wave network in a NL scenario is given by\vspace{-0.5em}
\begin{align}
\mathcal{P}^{\mu}_s(\{\nu_i\})=\sum\nolimits_{i=1}^{L} f_i \bigg\{1-\textup{exp}(-\tilde{k} \,\,p_{\mu_i} (\frac{\tilde{\eta}_i}{Q_i})^{\delta_\mu})\bigg\},\label{muwave}
\end{align}
where $\tilde{k}=\pi \lambda_\mu \Gamma(\delta_\mu+1), \tilde{\eta}_i=\frac{\mathrm{P}_\mu}{\sigma^2_{\mu_{x_i}}},\delta_\mu=\frac{2}{\alpha_\mu},Q_i = 2^{\rho_i} - 1, \rho_i${\footnote{By normalizing the whole bandwidth, $\rho_{\textit{max}}=1$ bits/s/Hz and $N_{x_i} \in [0,1]$ expresses the portion of the whole bandwidth. Hence, $\rho_i \leq 1$.}}$ = N_{x_i} \nu_i$.
\end{theorem1}
The optimization problem \textbf{P1} is now convex and the objective function can be rewritten as\vspace{-0.5em}
\begin{align}\label{2}
\mathcal{P}_s(\{\nu_i\}) &= \sum\nolimits_{i=1}^{L}f_i\Bigg[1 - \textup{exp}\bigg(-\sum\nolimits_{j\in\{\LOS,\NLOS\}}k_j p_{m_i} Z_j\big(\frac{\eta_i G_{x_i}}{Q_i }\big)-\hat{k}p_{m_i}\big(\frac{\eta_i G_{x_i}}{Q_i}\big)^{\delta_{m_\NLOS}}\bigg)\Bigg] p_{mw} \nonumber \\&+ \sum\nolimits_{i=1}^{L} f_i \bigg\{1-\textup{exp}(-\tilde{k} \,\,p_{\mu_i} (\frac{\tilde{\eta}_i}{Q_i})^{\delta_\mu})\bigg\} \,\, p_{\mu w}.
    \end{align}
    
The Lagrangian function of the optimization problem~\eqref{eq: objective function1}-\eqref{eq: lagrangian constraint3} is \vspace{-1.0em}
\begin{align}\label{eq:lagrange}
&\mathcal{L}\left(\{p_{m_i}\},\{p_{\mu_i}\},\tilde{\omega},\hat{\omega},\{\tilde{\mu}_i\},\{\hat{\mu}_i\}\right)\nonumber \\
&= \sum\nolimits_{i=1}^{L}f_i\Bigg[1 - \textup{exp}\bigg(-\sum\nolimits_{j\in\{\LOS,\NLOS\}}k_j p_{m_i} Z_j\big(\frac{\eta_i G_{x_i}}{Q_i }\big)-\hat{k}p_{m_i}\big(\frac{\eta_i G_{x_i}}{Q_i}\big)^{\delta_{m_\NLOS}}\bigg)\Bigg] p_{mw} \nonumber \\&+ \sum\nolimits_{i=1}^{L} f_i \bigg\{1-\textup{exp}(-\tilde{k} \,\,p_{\mu_i} (\frac{\tilde{\eta}_i}{Q_i})^{\delta_\mu})\bigg\} \,\, p_{\mu w}
-\tilde{\omega}\left(\sum\nolimits_{i=1}^{L} p_{m_i} -\cachesizemm\right)- \hat{\omega}\left(\sum\nolimits_{i=1}^{L} p_{\mu_i}-\cachesizemu\right)\nonumber \\
&-\sum\nolimits_{i=1}^{L} \tilde{\mu}_i(p_{m_i}-1)-\sum\nolimits_{i=1}^{L}\hat{\mu}_i(p_{\mu_i}-1),
\end{align}
where  $\tilde{\omega},\hat{\omega},\tilde{\mu}_i$ and $\hat{\mu}_i$ are the Lagrangian multipliers associated with the constraints~\eqref{eq: lagrangian constraint1}-\eqref{eq: lagrangian constraint3}, respectively. Since Slater's condition satisfies the optimization problem, the optimal solution of the problem can be achieved by solving its dual, which can be written as\vspace{-1.0em}
\begin{align}\label{dual_problem}\vspace{-0.5em}
\min_{\tilde{\omega},\hat{\omega},\{\tilde{\mu}_i\},\{\hat{\mu}_i\}\geq0}\max_{\{p_{m_i}\},\{p_{\mu_i}\}}\!\!\mathcal{L}\left(\{p_{m_i}\},\{p_{\mu_i}\},\tilde{\omega},\hat{\omega},\{\tilde{\mu}_i\},\{\hat{\mu}_i\}\right).
\end{align}
The dual problem in~\eqref{dual_problem} is then solved in an iterative fashion which alternates between a \textit{sub-problem}, updating the caching probability variables  $\{p_{m_i}\}$ and $\{p_{\mu_i}\}$ by fixing the Lagrangian multipliers $\left(\tilde{\omega},\hat{\omega},\{\tilde{\mu}_i\},\{\hat{\mu}_i\}\right)$, and a \textit{master problem}, computing new Lagrangian multipliers based on the obtained caching probabilities.
Further, in the sub-problem, by taking the partial derivative of \eqref{eq:lagrange} with respect to $p_{m_i}$, we can find the optimal caching probabilities for mmWave BSs. Accordingly,
\begin{align}\label{first order}
&\frac{\partial \mathcal{L}(\{p_{m_i}\}, \tilde{\omega}, \{\tilde{\mu}_i\})}{\partial p_{m_i}}=f_i \textup{exp}[p_{m_i} (-A_i-B_i)] + \tilde{\omega} + \tilde{\mu}_i
\end{align}
where 
$A_i = \sum_{j\in\{\LOS,\NLOS\}}k_j Z_j(\frac{\eta_i G_{x_i}}{Q_i})$ and $B_i = \hat{k} (\frac{\eta_i G_{x_i}}{Q_i})^{\delta_{m_\NLOS}}.$
The optimal caching probability is now given by
\begin{align}
p_{m_i} = \left[\frac{1}{A_i+B_i}\textrm{log}\bigg(\frac{f_i p_{mw}}{\tilde{\omega} +\tilde{\mu}_i}\bigg)\right]^+\label{optimal_mm},
\end{align}
where $[x]^+ = \textup{max}\{0,x\}$.
Likewise, the optimal caching probability for $\mu$Wave BS is given by
\begin{align}
p_{\mu_i}=\bigg[\frac{1}{\hat{T}_i \tilde{k}} \textrm{log}\bigg(\frac{p_{\mu w}f_i}{\hat{\omega} +\hat{\mu}_i}\bigg)\bigg]^+\,\label{optimal_mu},
\end{align}
where  $\hat{T}_i = (\frac{\tilde{\eta}_i}{Q_i}\big)^{\delta_\mu}$.
The Lagrange multipliers are updated using subgradient method. The algorithm to find the optimal caching probabilities $\{p^*_{m_i}\}$ and $\{p^*_{\mu_i}\}$ is given in Algorithm \ref{algorithm111}.
\begin{algorithm} [t!]\small{
\caption{\textbf{: Computation of $\{p_{m_i}^*\}$ and $\{p_{\mu_i}^*\}$}}
\label{algorithm111}
\renewcommand{\arraystretch}{0.5}
\begin{algorithmic}[1]
\STATE$\text{Initialize}:\,\tilde{\omega},\hat{\omega},\{\tilde{\mu}_i\},\{\hat{\mu}_i\}$\\
\STATE$\text{Compute}\,\, p_{m_i}$ using~\eqref{optimal_mm}.
\STATE$\text{Compute} \,\, p_{\mu_i}$ using~\eqref{optimal_mu}.
\STATE$\text{Update the Langrangian multipliers}\;\tilde{\omega},\hat{\omega},\{\tilde{\mu}_i\}, \text{and}\;\{\hat{\mu}_i\}$.
\STATE$\text{Repeat steps} \; 2-4$\; until convergence.
\STATE$p^*_{m_i}\leftarrow p_{m_i}(\tilde{\omega}^*,\tilde{\mu}^*_i), p^*_{\mu_i}\leftarrow p_{\mu_i}(\hat{\omega}^*,\hat{\mu}^*_i)$.
\end{algorithmic}}\normalsize
\end{algorithm}
\vspace{-1.0em}
\subsection{Optimal caching probabilities under IL scenario}
Compared to the NL case, the interference part in \eqref{eq: received_sinr} and \eqref{eq: received_sinr_mu} will be dominant when the number of mm/$\mu$Wave BSs (\emph{i.e.,} density) increases and blockage density in the network decreases. 
In this subsection, the ASP of file delivery under a IL case is given.
\begin{theorem1}
The IL ASP of file delivery in a mmWave network is given by
    \begin{align}
\mathcal{P}^{mm}_s(\nu_i) \leq&\sum\nolimits_{i=1}^{\textit{L}} \left\{f_i \left\{\sum\nolimits_{l=1}^{\hat{m}} \binom{\hat{m}}{l} (-1)^{l+1} \sum\nolimits_{j\in\{\LOS,\NLOS\}}  p_j 
 \right.\right.\nonumber \\ 
&\left.\left.\left\{\prod_{\hat{q}=1}^{3}\prod_{\hat{j} \in \{\LOS,\NLOS\}}  \textup{exp}\Big[-\overline{\lambda}_{m_i} \int_{0}^{\infty}\Big(1- \Big(\frac{1}{1+\frac{A l Q_i  \hat{G}_{\hat{q}}  r^{-\alpha_{\hat{j}}}_t}{G_{x_i} r^{-\alpha_j}_{x_i} \hat{m}}}\Big)^{\hat{m}}\Big) p_{\hat{j}} 2 \pi p_{\hat{q}} \, r_t  dr_t\Big]\right\}\right\}\right\}\,.
\end{align}
\end{theorem1}
\begin{proof}
This proof can be obtained similar to Theorem \ref{theorem mmwave} with interference part only.
\end{proof}
\begin{theorem1}
The IL ASP of file delivery in a $\mu$Wave network is given similarly as
    \begin{align}
\mathcal{P}^{\mu}_s(\nu_i)\leq&\sum\nolimits_{i=1}^{\textit{L}}f_i\left\{\sum\nolimits_{l=1}^{\hat{m}}\left\{\binom{\hat{m}}{l}(-1)^{l+1} \right.\right.\nonumber \\ 
& \left.\left.\textup{exp}\Big[-2\pi \overline{\lambda}_{\mu_i}\int_{0}^{\infty} r_t\bigg(1-\big(1+\frac{l A Q_i  r^{-\alpha_\mu}_t}{r^{-\alpha_\mu}_{x_i}\hat{m}}\big)^{-\hat{m}}\bigg)dr_t\Big]\right\}\right\}\,,
\end{align}
where all the parameters are as defined before.
\end{theorem1}
\begin{proof}
This proof can be obtained by leveraging on the derivation of Theorem \ref{theorem_mu}.
\end{proof}
After establishing the ASP of file delivery for the hybrid network in an IL scenario, the objective function of the optimization problem \textbf{P1} for this IL case can now be rewritten as in~\eqref{ASP_IL_Scenario}, shown on the top of the next page.
\begin{figure*}
\begin{align}\label{ASP_IL_Scenario}
        &\mathcal{P}_s(\{\mu_i\})
        \leq \sum\nolimits_{i=1}^{\textit{L}} \left\{f_i \left\{\sum\nolimits_{l=1}^{\hat{m}} \binom{\hat{m}}{l} (-1)^{l+1} \sum\nolimits_{j\in\{\LOS,\NLOS\}}  p_j 
 \right.\right.\nonumber\\ 
&\left.\left.\left\{\prod_{\hat{q}=1}^{3}\prod_{\hat{j} \in \{\LOS,\NLOS\}}  \textup{exp}\Big[-\overline{\lambda}_{m_i} \int_{0}^{\infty}\Big(1- \Big(\frac{1}{1+\frac{A l Q_i  \hat{G}_{\hat{q}}  r^{-\alpha_{\hat{j}}}_t}{G_{x_i} r^{-\alpha_j}_{x_i} \hat{m}}}\Big)^{\hat{m}}\Big) p_{\hat{j}} 2 \pi p_{\hat{q}} \, r_t  dr_t\Big]\right\}\right\}\right\} \,\, p_{mw} \nonumber \\
&+\hspace{-0.20em}\sum\nolimits_{i=1}^{\textit{L}}f_i\left\{\sum_{l=1}^{\hat{m}}\left\{\binom{\hat{m}}{l}(-1)^{l+1}\hspace{-0.20em}\textup{exp}\Big[-2\pi \overline{\lambda}_{\mu_i}\int_{0}^{\infty} r_t\bigg(1-\big(1+\frac{l A Q_i  r^{-\alpha_\mu}_t}{r^{-\alpha_\mu}_{x_i}\hat{m}}\big)^{-\hat{m}}\bigg)dr_t\Big]\right\}\right\}\,p_{\mu w}\nonumber\\
    &= \sum\nolimits_{i=1}^{L} f_i \sum\nolimits_{l=1}^{\hat{m}}\binom{\hat{m}}{l} (-1)^{(l+1)} \sum\nolimits_{j\in\{\LOS,\NLOS\}} p_j \nonumber \\
    &\times \textrm{exp}\Bigg[-(1-{ p_{m_i}}) \lambda_m \underbrace{\sum_{\hat{q}=1}^{3}\sum\nolimits_{\hat{j}\in\{\LOS,\NLOS\}} \int_{0}^{\infty}\Big(1- \Big(\frac{1}{1+\frac{A l Q_i  \hat{G}_{\hat{q}}  r^{-\alpha_{\hat{j}}}_t}{G_{x_i} r^{-\alpha_j}_{x_i} \hat{m}}}\Big)^{\hat{m}}\Big) p_{\hat{j}} 2 \pi p_{\hat{q}} \, r_t  dr_t}_{\hat{Z}(i,l)}\Bigg] p_{mw}\nonumber \\
    &+\hspace{-0.20em} \sum_{i=1}^{L} f_i \sum_{l=1}^{\hat{m}}\hspace{-0.10em}\binom{\hat{m}}{l} (-1)^{(l+1)}\times \textrm{exp}\hspace{-0.15em} \Big[\hspace{-0.1em}-(1-{ p_{\mu_i}})\lambda_\mu\underbrace{\int_{0}^{\infty} r_t\bigg(1-\big(1+\frac{l A Q_i  r^{-\alpha_\mu}_t}{r^{-\alpha_\mu}_{x_i}\hat{m}}\big)^{-\hat{m}}\bigg)2\pi dr_t}_{\hat{W}(i,l)} \Big] p_{\mu w}\nonumber\\
    =&\underbrace{\sum\nolimits_{i=1}^{L} f_i\sum\nolimits_{l=\textrm{odd number}} \binom{\hat{m}}{l} \sum\nolimits_{j\in\{\LOS,\NLOS\}} p_j \textrm{exp}[-\lambda_m \hat{Z}(i,l)]\textup{exp}[{p_{m_i}} (\lambda_m\hat{Z}(i,l) ]p_{mw}}_{h({\{p_{m_i}\}})} \nonumber \\&-\underbrace{\sum\nolimits_{i=1}^{L} f_i\sum\nolimits_{l=\textrm{even number}} \binom{\hat{m}}{l} \sum\nolimits_{j\in\{\LOS,\NLOS\}} p_j \textrm{exp}[-\lambda_m \hat{Z}(i,l)]\textup{exp}[{p_{m_i}} (\lambda_m\hat{Z}(i,l) ]p_{mw}}_{g({\{p_{m}\}})} \nonumber \\
    &+\underbrace{\sum\nolimits_{i=1}^{L} f_i \sum\nolimits_{l=\textrm{odd number}}\binom{\hat{m}}{l} \textrm{exp}(-\lambda_\mu \hat{W}(i,l))\textrm{exp}[{p_{\mu_i}} (\lambda_\mu \hat{W}(i,l)  )]p_{\mu w} }_{hh({ \{p_{\mu_i}\}})} \nonumber \\ &- \underbrace{\sum\nolimits_{i=1}^{L} f_i \sum\nolimits_{l=\textrm{even number}}\binom{\hat{m}}{l} \textrm{exp}(-\lambda_\mu \hat{W}(i,l))\textrm{exp}[{p_{\mu_i}} (\lambda_\mu \hat{W}(i,l)  )]p_{\mu w} }_{gg({\{p_{\mu_i}\}})}
    \end{align}
    \hrule\vspace{-1.5em}
    \end{figure*}
    In the above, $h({\{p_{m_i}\}}), g({\{p_{m_i}\}}), hh({\{p_{\mu_i}\}}), gg({ \{p_{\mu_i}\}})$ are convex. 
    %
    The optimization problem to find the optimal file placement scheme can now be formulated as a standard {\it DCP problem} and given as\vspace{-1.0em}
        \begin{align}
    \mathbf{P2}: \,\,\,\,\,\min_{\{p_{m_i}\},\{p_{\mu_i}\}} & \left[-h(\{p_{m_i}\}) + g(\{p_{m_i}\}) - hh(\{p_{\mu_i}\})+gg(\{p_{\mu_i}\})\right]     \label{eq40}\\
    \text{s.t.}\quad& \sum\nolimits_{i=1}^{L}p_{m_i} \leq \cachesizemm,\\
    & \sum\nolimits_{i=1}^{L} p_{\mu_i} \leq \cachesizemu,\\
    & 0\leq p_{m_i}\leq 1,\;\;\text{\&}\;\;0\leq p_{\mu_i} \leq 1\,,\quad \forall i \in \mathcal{S}.
    \end{align}Now, it can be noted that since mmWave and $\mu$Wave networks can be independently represented through their respective association probabilities, we can separately calculate the optimal caching probabilities for mmWave BSs and $\mu$Wave BSs by rewriting the above optimization problem into two sub-DCP problems. Based on \cite{shen2016disciplined}, we propose an iterative algorithm as given in Algorithm \ref{algorithm11} to obtain the optimal caching probabilities for both mmWave and $\mu$Wave networks by separately converting the sub-DC objective functions to convex functions. 
     Below, we give the proof of convergence of Algorithm \ref{algorithm11}.
   \begin{algorithm} [t!]\small{
\caption{\textbf{: Computation of $\{p_{m_i}^*\}$ and $\{p_{\mu_i}^*\}$}}
\label{algorithm}\label{algorithm11}
\begin{algorithmic}[1]
\STATE $\textbf{Initialize}:\,\text{counter}\,\,k=0, \{p^{0}_{m_i}\}, \{p^{0}_{\mu_i}\},\,\,\text{step size }\,\Delta = 10^{-4} \,\text{ and threshold } \hat{\delta} = 10^{-5}$\\
\STATE $\textbf{Repeat}$\\
\STATE $\hspace{1.5cm}\textbf{Compute:}  \,\, \hat{h}(\textbf{p}^{k+1}_{m};\textbf{p}^{k}_{m}) = h(\mathbf{p}^k_{m}) + \nabla \mathbf{h}^{\textrm{T}}(\mathbf{p}^{k}_{m}) (\mathbf{p}^{k+1}_{m} - \mathbf{p}^{k}_{m})$\\
$\hspace{3.1cm}\hat{hh}(\mathbf{p}^{k+1}_{\mu};\mathbf{p}^{k}_{\mu})=hh(\mathbf{p}^{k}_\mu) + \nabla \mathbf{hh}^\textrm{T}(\mathbf{p}^{k}_\mu) \, (\mathbf{p}^{k+1}_\mu - \mathbf{p}^{k}_\mu)$\\
\STATE $\hspace{1.5cm}\textbf{Solve 1:}\,\,\text{Set the value of }\{p^{k+1}_{m_i}\}\,\,\text{to be a solution of }$\\
$\hspace{3.2cm}\text{minimize} \,\, g(\{p^{k+1}_{m_i}\}) - \hat{h}(\{p^{k+1}_{m_i}\};\{p^{k}_{m_i}\})$\\
$\hspace{3.2cm}\text{subject to}\,\, \sum_{i=1}^{L} \, p^{k+1}_{m_i} \leq M,\,\, 0\leq p^{k+1}_{m_i} \leq 1, \, \forall{i} \in \mathcal{S}$\\
\STATE$\hspace{1.5cm}\textbf{Solve 2:}\,\,\text{Set the value of }\{p^{k+1}_{\mu_i}\}\,\,\text{to be a solution of }$\\
$\hspace{3.2cm}\text{minimize} \,\, gg(\{p^{k+1}_{\mu_i}\}) - \hat{hh}(\{p^{k+1}_{\mu_i}\};\{p^{k}_{\mu_i}\})$\\
$\hspace{3.2cm}\text{subject to}\,\, \sum_{i=1}^{L} \, p^{k+1}_{\mu_i} \leq N,\,\, 0\leq p^{k+1}_{\mu_i} \leq 1, \, \forall{i} \in \mathcal{S}$\\
\STATE $\hspace{1.5cm}\textbf{Update:} \,\,\, k = k + 1.\,\,$
\STATE 
$\hspace{3.2cm} \mathbf{p}^{k}_{m}=\mathbf{p}^{k-1}_{m}+sign(\mathbf{p}^{k}_{m}-\mathbf{p}^{k-1}_{m})\times\nabla \mathbf{h}^{\textrm{T}}(\mathbf{p}^{k-1}_{m}) \, \times\Delta,$ \\
$\hspace{3.2cm} \mathbf{p}^{k}_{\mu}=\mathbf{p}^{k-1}_{\mu}+sign(\mathbf{p}^{k}_{\mu}-\mathbf{p}^{k-1}_{\mu})\times\nabla \mathbf{h}^{\textrm{T}}(\mathbf{p}^{k-1}_{\mu}) \, \times\Delta$, \\
\hspace{3.2cm}$\text{where}\,\,
	sign(x)= \left\{ 
	\begin{array} {l l}
	1,\,\,\,\,\, x> 0\\
	0,\,\,\,\,\, x=0\\
	-1, \, x< 0
	\end{array} \right.
$\\
\STATE $\textbf{Until }\,\, \text{convergence or maximum iteration number is reached.}$\\
\end{algorithmic}}\normalsize
\end{algorithm}\vspace{-1.0em}
\begin{theorem1}\label{Theorem_DC}
The original objective function  in \eqref{eq40} of the DCP problem \textbf{P2} can be convexified by replacing it with its upper bound. The proposed algorithm is then convergent with respect to an increase in the iteration number.
\end{theorem1}
    \begin{proof}
    Let $\mathbf{p}^k_m$ and $\mathbf{p}^k_\mu$ be the feasible points for problem \textbf{P2}. Applying Taylor series approximation on $h(\{p^{k+1}_{m_i}\})$ and $hh(\{p^{k+1}_{\mu_i}\})$ at feasible points $\mathbf{p}^k_{m}$ and $\mathbf{p}^k_{\mu}$, the objective function is rewritten as\vspace{-1em}
     \begin{equation}\label{aaa111}
     v_{k+1} = g(\mathbf{p}^{k+1}_m) - h(\mathbf{p}^{k+1}_m) + gg(\mathbf{p}^{k+1}_\mu) - hh(\mathbf{p}^{k+1}_\mu),\vspace{-0.5em}
     \end{equation} 
     such that \eqref{aaa111}  is now convex. Since the region of feasible solution remains the same, the feasible points $\mathbf{p}^k_m$ and $\mathbf{p}^k_\mu$ are also feasible for the convexified problem and other feasible points $\mathbf{p}^{k+1}_m$ and $\mathbf{p}^{k+1}_\mu$ that exist for the convexified problem are also the feasible points of the problem \textbf{P2}. Furthermore, for all $\mathbf{p}_m$ and $\mathbf{p}_\mu$ using Taylor series approximation, the convexity of $h$ and $hh$ gives us \vspace{-1.75em}
     \begin{equation}
     \hat{h}(\mathbf{p}^{k+1}_m;\mathbf{p}^k_m) \leq h(\mathbf{p}^{k+1}_m),\vspace{-0.5em}
     \end{equation}
     and\vspace{-1.5em}
     \begin{equation}
     \hat{hh}(\mathbf{p}^{k+1}_\mu;\mathbf{p}^k_\mu) \leq hh(\mathbf{p}^{k+1}_\mu).\vspace{-0.75em}
     \end{equation}
      Hence, if  $\mathbf{p}^0_m$ and $\mathbf{p}^0_\mu$ are chosen to be feasible, all corresponding iterates will be feasible. 
    Now, we show that the objective value converges over the iterations. According to the inequalities \vspace{-1.0em}
    \begin{eqnarray}
    g(\mathbf{p}_m) < h(\mathbf{p}_m),\label{inequality1}\\
    gg(\mathbf{p}_\mu) < hh(\mathbf{p}_\mu),\label{inequality2}
    \end{eqnarray}
    \begin{equation}
   \text{and}\,\,\, v_{k+1} \leq \underbrace{g(\mathbf{p}^{k+1}_m) - \hat{h}(\mathbf{p}^{k+1}_m;\mathbf{p}^{k}_m) + gg(\mathbf{p}^{k+1}_\mu) - \hat{hh}(\mathbf{p}^{k+1}_\mu;\mathbf{p}^{k}_\mu)}_{\hat{v}_{k+1}},\label{inequality3}
    \end{equation}
we minimize the value of $\hat{v}_{k+1}$ at each iteration $k$, and obtain $v_k$ by using previous $\mathbf{p}^k_m$ and $\mathbf{p}^k_\mu$ such that\vspace{-1.50em}
\begin{equation}
v_k \geq \hat{v}_{k+1} \geq v_{k+1}.\vspace{-0.5em}
\end{equation}    
The value of the above objective function is now non-increasing and will always converge, possibly to negative infinity, which concludes the proof.
\end{proof}
\begin{table}[t!]
\footnotesize
\renewcommand{\arraystretch}{0.65}
\centering
\caption{Parameter values}\vspace{-1em}
\label{table1}
\begin{tabular}{|l|l|l|}

\hline
{\textbf{Parameter notation}}&{\textbf{Physical meaning}}&{\textbf{Values}}\\  \hline
{$\eta_i=\frac{\mathrm{P}_m}{\sigma^2_{m_{x_i}}}, \forall i \in \mathcal{S}$}&{SNR of the typical user for file $i$ from the mmWave serving BS}&{54 (dB)}\\ \hline
{$\tilde{\eta}_i = \frac{\mathrm{P}_\mu}{\sigma^{2}_{\mu_{x_i}}}, \forall i \in \mathcal{S}$}&{SNR of the typical user for file $i$ from the $\mu$Wave serving BS}&{104 (dB)}\\ \hline
{$\theta$} & {Mainlobe beamwidth} & {$\pi/6$} \\ \hline
{$G^M$ / $G^m$} & {Mainlobe antenna gain / sidelobe antenna gain} & {15 (dB) / -15 (dB)} \\ \hline
{$G_{x_i}$ }  & {Effective antenna gain between the serving mmWave BS and typical user}  &  {225 (dB)}  \\ \hline
 {$\alpha_\LOS$ / $\alpha_\NLOS$} & {Path loss exponent of LoS and NLoS} &{2 / 4} \\ \hline
 {$\alpha_\mu$}&{$\mu$Wave path loss exponent}&{3.5}\\ \hline
 {$\lambda_m$}&{mmWave BS density}&{5$\times 10^-5$ (nodes/m$^2$)}\\ \hline
 {$\lambda_\mu$}&{$\mu$Wave BS density}&{$10^-6$ (nodes/m$^2$)}\\ \hline
 {$\beta$}&{Blockage density}&{0.008}\\ \hline
 {$\upsilon$}&{Skewness of the content popularity}&{0.8}\\ \hline
 {$L$}&{The number of files}&{10}\\ \hline
 {$R$}&{The radius of the bounded coverage region}&{500 (m)}\\ \hline
 {$\cachesizemm / \cachesizemu$}&{Cache size of mmWave/$\mu$Wave BS}&{5 / 6}\\ \hline
 {$\hat{m}$}&{Nakagami fading parameter for mmWave ($\mu$Wave) channel}&{10 (1)}\\ \hline
 {$\rho_{\textrm{max}}$}&{The maximum file delivery rate}&{1 (bit/sec/Hz)}\\ \hline
 {$\rho_i, \forall i \in \mathcal{S}$}&{The rate for $i$-th file delivery (\emph{i.e.,}$\rho_i \in [0,\rho_{\textrm{max}}]$)}&{0.8 (bits/sec/Hz)}\\ \hline
\end{tabular}
\vspace{-1.5em}
\end{table}
\vspace{-1.5em}
\section{Numerical results}\vspace{-.5em}
After developing the analytical framework in the previous sections, we now evaluate the performance (ASP of file delivery) of the proposed caching placement strategy with respect to Algorithm \ref{algorithm111} and \ref{algorithm11}. Unless otherwise stated, most of the parameters used and their corresponding values are inspired from literature and given in Table \ref{table1}. A uniform target rate for each file is considered for simplicity throughout the analysis.

We begin by evaluating the optimal caching probabilities in a NL hybrid network for varying densities of mmWave and $\mu$Wave networks and different Nakagami parameters in Fig. \ref{fig3}. It is worth noting that according to the CDF of the process $\{\frac{r^{\alpha}}{\mathcal{X}}\}$ in \eqref{eq11}, when $\lambda_m p_{m_i}$ becomes higher, the minimum of the CDF of $\{\frac{r^{\alpha}}{\mathcal{X}}\}$ increases. Accordingly, the ASP of file delivery will be higher due to the increase in intensity measure. In fact, the reciprocal of  $\{\frac{r^{\alpha}}{\mathcal{X}}\}$ represents the effective channel gain and hence, with the increase in density, the probability to obtain higher channel gain also increases. 
However, it can also be seen that when the density increases, the caching probabilities for the most popular files decrease and tend to be uniformly distributed. 
%
This means that by sacrificing higher channel gain for a few specific contents, we can increase the hitting probability of all contents (\textit{content diversity gain}) such that the optimal ASP of file delivery can be achieved. 
%
Therefore, there is a tradeoff between channel gain and cache hit. 
\begin{figure}[t]
\centering
\includegraphics[width=0.65\linewidth]{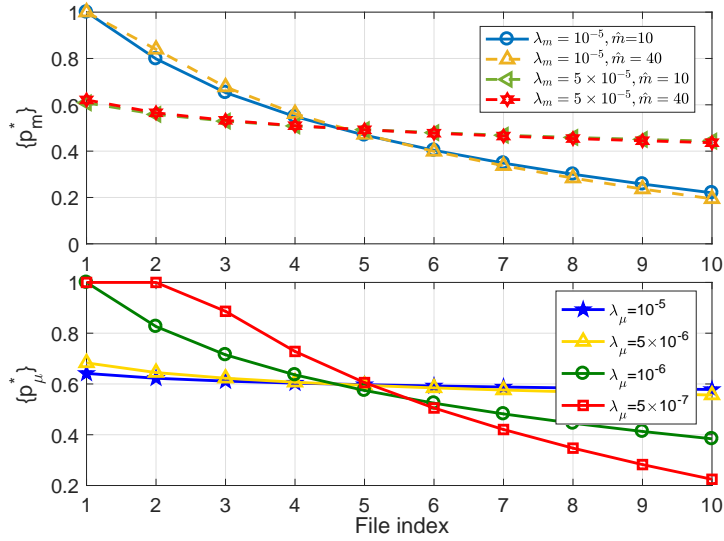}\vspace{-1.25em}
\caption{Optimal caching probability v.s. varying $\lambda$ and $\hat m$ values under NL scenario.}
\label{fig3}\vspace{-1.5em}
\end{figure}
Besides, in the same figure, we also evaluate the effect of Nakagami fading parameter $\hat{m}$, which relates to the channel power gain. The figure shows that the proposed algorithm is not affected by $\hat{m}$, {which shows that BS density is a more significant parameter than the fading parameter $\hat{m}$}. The above explanation holds true for $\mu$Wave systems as well as can be seen from the tradeoff in performance for $\mu$Wave optimal caching probabilities {in terms of BS density}. 

We now compare the ASP of file delivery of the proposed content placement strategy as shown in Algorithm \ref{algorithm111} in a NL hybrid network with three different content placement strategies: 1) caching the $M$ most popular contents (MC), 2) caching the contents uniformly (UC), and 3) caching contents randomly (RC) in Fig. \ref{fig1}.
It is evident from the figure that the proposed caching placement scheme is superior to the MC, UC and RC in terms of ASP of file delivery for varying content popularities. When the skewness $\upsilon$ of the content popularity distribution is close to zero, meaning of which is that the content popularity is uniformly distributed and uniformly requested by users, the proposed caching placement strategy is significantly better than the others. On the other hand, while MC  is better than RC for higher values of $\upsilon$ and vice versa for low $\upsilon$, both perform comparatively inferior to UC throughout almost the entire range of $\upsilon$.  
\begin{figure*}[t!]
\minipage{0.52\textwidth}
  \includegraphics[width=\linewidth]{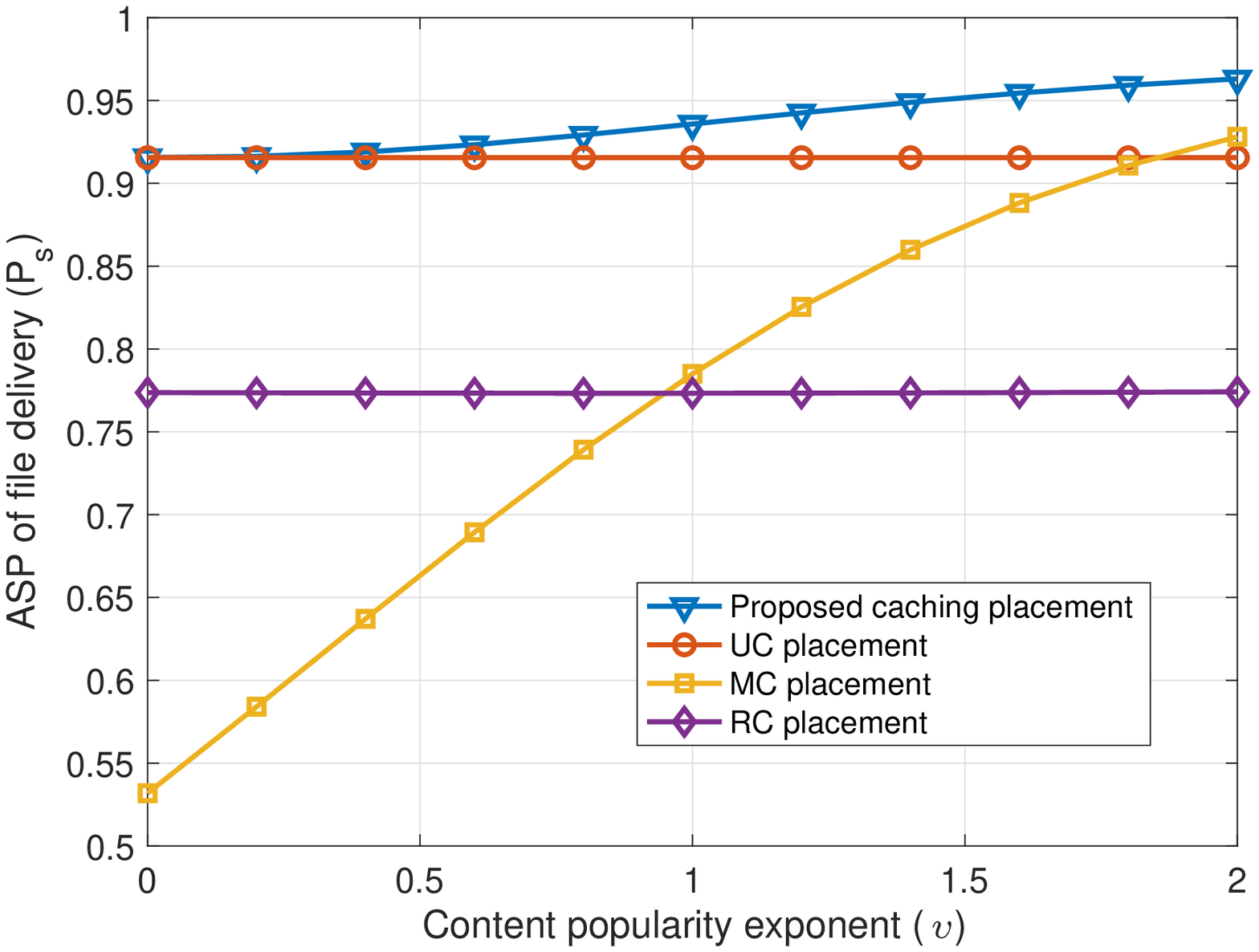}
\caption{ASP of file delivery for various caching placement \\strategies under NL scenario.}
\label{fig1}
\endminipage\hfill
\minipage{0.52\textwidth}
  \includegraphics[width=\linewidth]{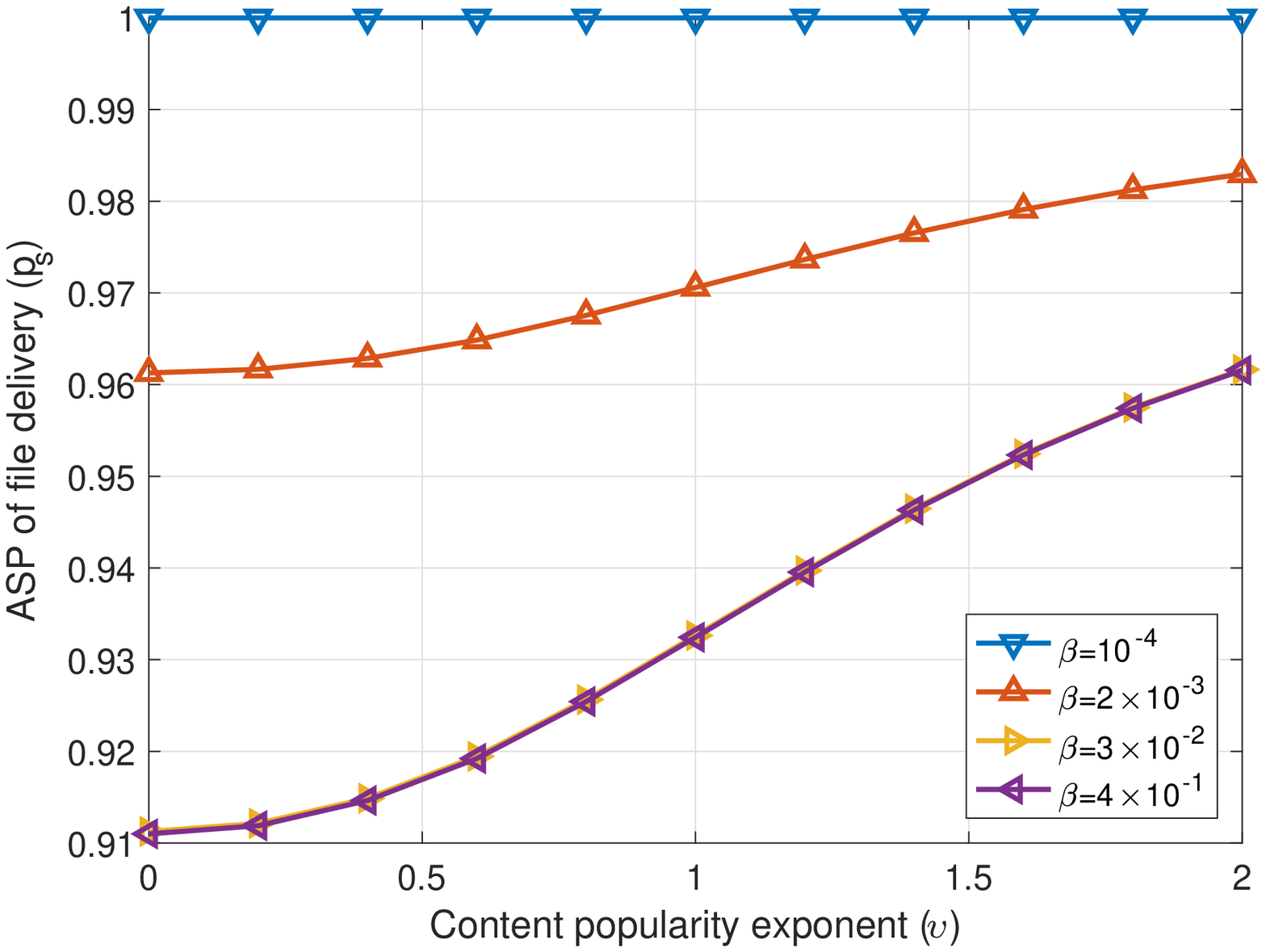}
\caption{ASP of file delivery for various blockage densities \\under NL scenario.}
\label{fig2}
\endminipage
\vspace{-2.50em}
\end{figure*}
Next, in Fig. \ref{fig2}, we evaluate the effect of blockages on our proposed algorithm for the NL scenario. 
It can be seen that when blockage density increases, the ASP of file delivery decreases. This is due to the fact that increasing blockages in the mmWave network results in the attenuation of the received signal. However, the decrease in optimal ASP is not very significant. This can be explained as: 1) for a substantial blockage density, with the increase in skewness, the number of files with higher probability requests decreases up to even less than the cache size of $\mu$Wave BSs, 2) the proposed algorithm makes sure that the higher caching probability of the most requested files is stored in the limited number of $\mu$Wave BSs, which are not affected by blockages, and 3) for the mmWave network, the algorithm also makes sure that the most requested files are stored only in the mmWave BSs with less average probability of NLOS than LOS. Hence, it can be concluded that the proposed algorithm is a {\it blockage-aware} optimal caching strategy.

\begin{figure}[t!]
\centering
\includegraphics[width=0.53\linewidth]{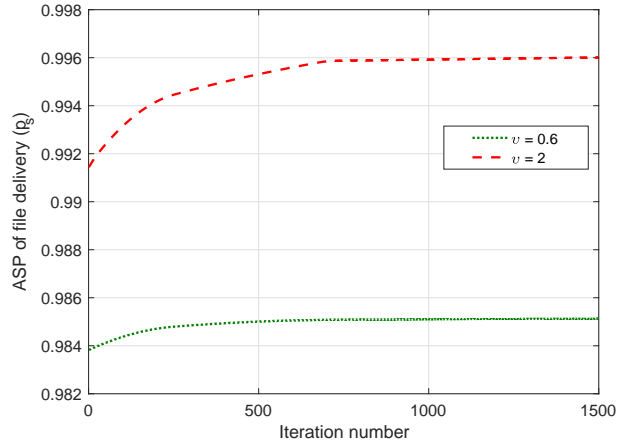}\vspace{-0.85em}
\caption{Convergence of Algorithm 2 under IL scenario for content popularity exponents $\upsilon = 0.6$ and $2$.}
\label{convergence}\vspace{-1.35em}
\end{figure}
\begin{figure*}[t!]
\minipage{0.52\textwidth}
  \includegraphics[width=\linewidth]{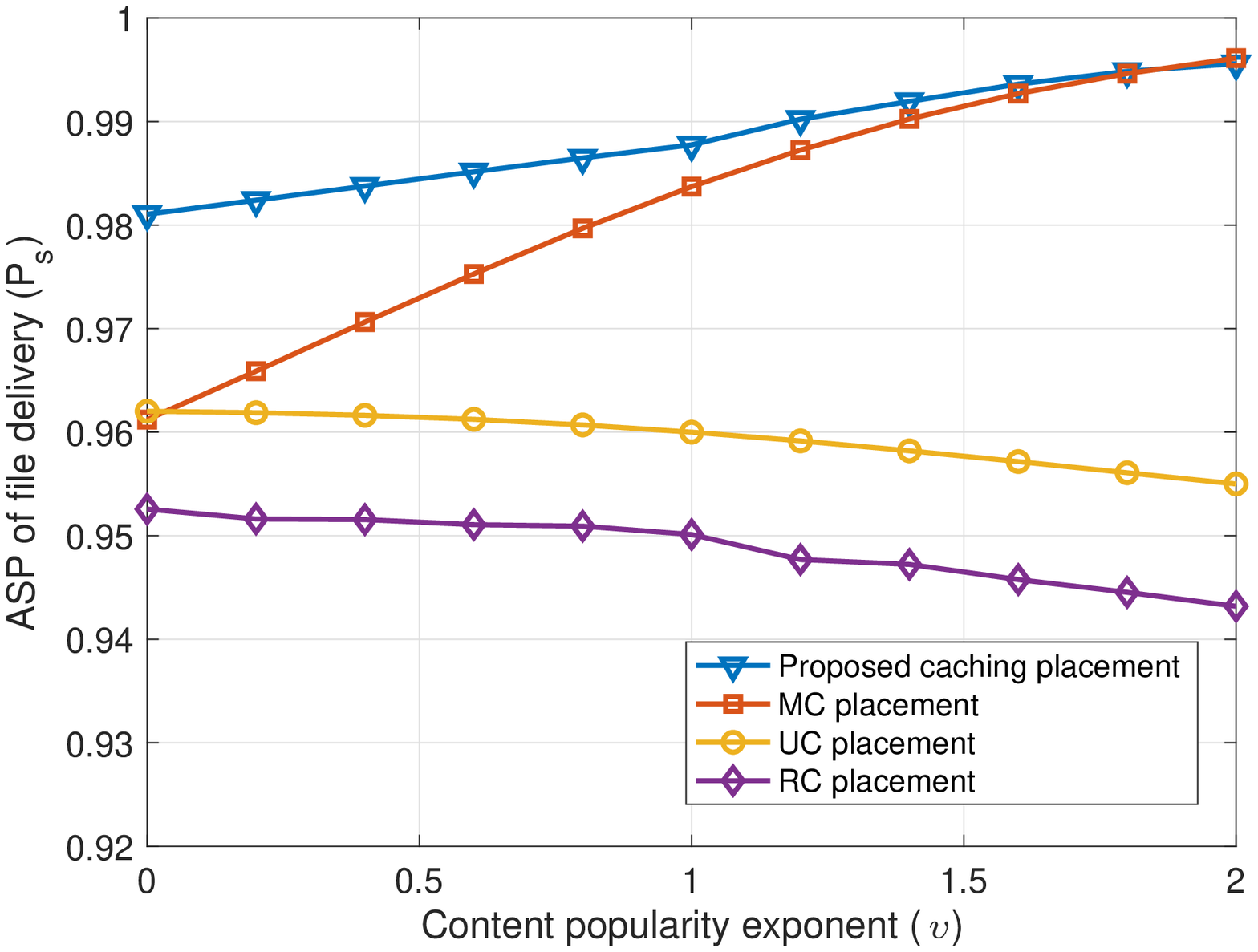}
  \vspace{-0.25em}
\caption{ASP of file delivery for various caching placement \\strategies under IL scenario.}
\label{fig4}
\endminipage\hfill
\minipage{0.52\textwidth}
  \includegraphics[width=\linewidth]{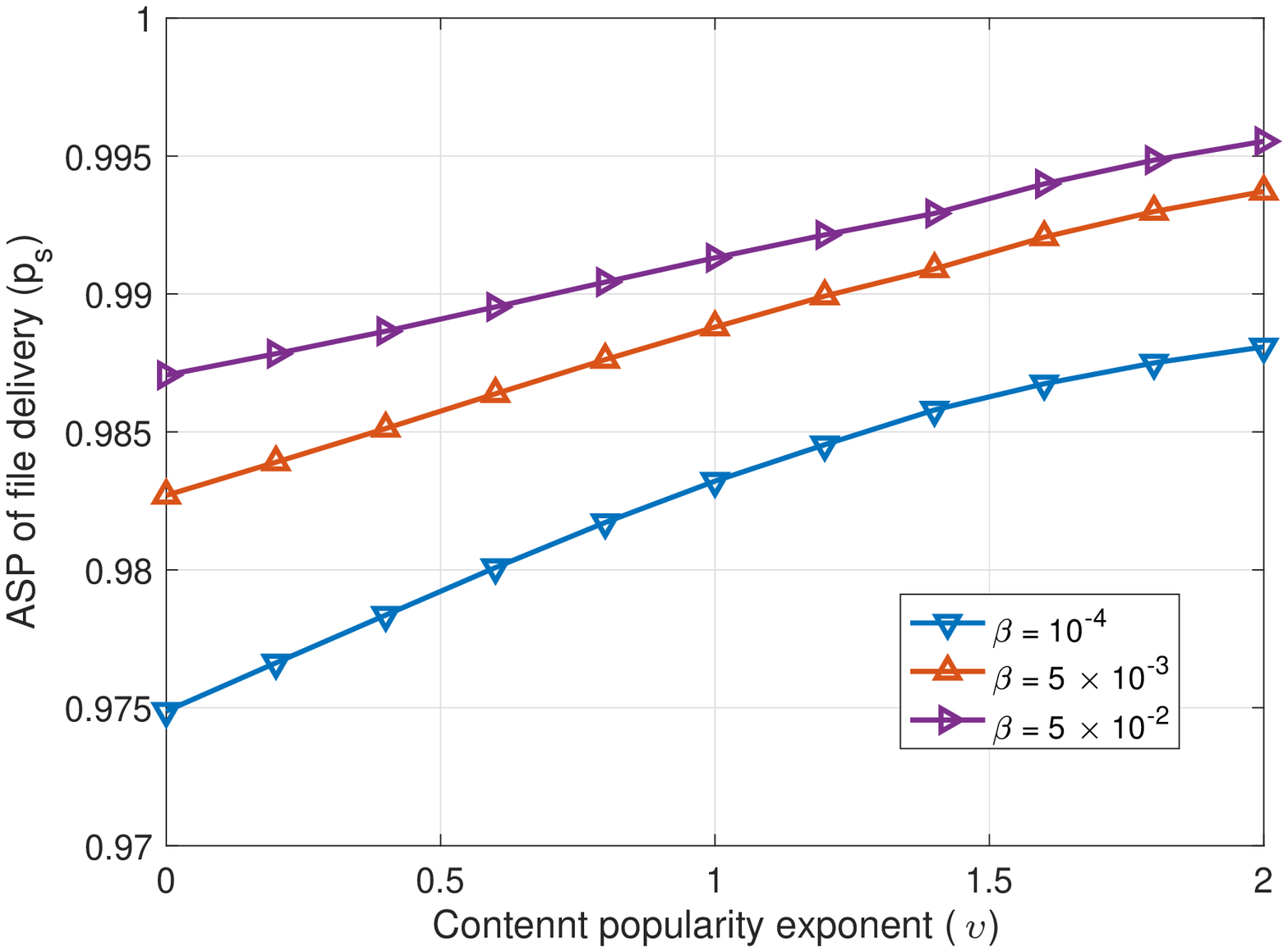}\vspace{-0.25em}
\caption{ASP of file delivery for various blockage densities\\ under IL scenario.}
\label{fig6}
\endminipage
\vspace{-3.0em}
\end{figure*}
After validating the results for the NL scenario, we now do the same for the IL case with similar parameter settings as in Table \ref{table1}, except for $\mu$Wave BS density $\lambda_\mu = 10^{-5}$, blockage density $\beta = 0.005$, and target rate $\rho_i = 0.08$.
 
 We begin by showing the evolution of Algorithm 2 in Fig. \ref{convergence} for $\upsilon = \{0.6, 2\}$. The monotonic increase of the cost function (ASP of file delivery) verifies the convergence of the proposed algorithm, which was also verified in Theorem \ref{Theorem_DC}.
 
Similar to the NL case, in Fig. \ref{fig4}, we show the superiority of the proposed caching placement scheme in IL scenario over three different content placement strategies: MC, UC, and RC. Like before, the proposed caching placement scheme is mostly superior to the MC, UC and RC in terms of ASP of file delivery for varying content popularities. When the skewness ($\upsilon$) of the content popularity distribution is close to zero, the proposed caching placement strategy is distinctively better than the others. However, at higher values of $\upsilon$, performance of MC is comparable to or slightly better than the proposed algorithm. This is due to the fact that higher values of $\upsilon$ means that the content popularity is not uniform and the contents are not uniformly requested by the users. Instead, only the most popular files are requested by the users which gives a performance edge to MC. Furthermore, while MC  is better than RC and UC for higher values of $\upsilon$ and close to UC for low $\upsilon$, UC performs better than RC throughout the entire range of $\upsilon$.  

Finally, in Fig. \ref{fig6}, we evaluate the effect of blockages on our proposed algorithm for the IL scenario. 
Unlike the NL case, when blockage density increases, the ASP of file delivery increases in the IL case.  The increase in optimal ASP is due to the effect of interference mitigation through blockages. More blockages in the network help in attenuating the interfering signals, which enhances the rate of file delivery. This can be considered as one of the very few instances when blockages are beneficial.
\vspace{-1em}
\section{Conclusion}\vspace{-.1em}
The coexistence of a hybrid mmWave and $\mu$Wave networks from the perspective of \emph{wireless caching} was shown.
Using stochastic geometric framework, 
that accounts for  not only interference among the operating base stations (BSs), uncertainties both in node locations and channel fading, path loss and loads at cache-enabled BSs, but also blockages in the mmWave network, we studied  the optimal probabilistic file caching placement at desirable mmWave/$\mu$Wave BSs.
In this regard, expressions for the association probability of a user to the mmWave or $\mu$Wave network was calculated, which was then used to derive closed form expression for average success probability (ASP) of file delivery in a noise-limited scenario and a upper bound for the same in a interference-limited scenario. Using the ASP of file delivery, we formulated the probabilistic content caching placement scheme as an optimization problem with respect to caching probabilities and accordingly proposed two algorithms, one each for noise-limited and interference-limited scenarios to acquire optimal caching probabilities by maximizing the ASP of file delivery.
Numerical evaluations were done with regards to the performance of the proposed content placement scheme under several essential factors, which demonstrated the superiority of the proposed caching scheme over others, even though certain trade-offs were observed. 
\vspace{-.5em}
\appendices
\vspace{-.5em}
\section{Proof of Theorem 2}\vspace{-.5em}
{To proof this theorem, we substitute \eqref{eq: received_sinr} and \eqref{eq: downlink_rate} into \eqref{eq: success_probability}. Accordingly, the conditional ASP of file delivery by the mmWave BSs can be reduced to
\begin{align}
&\mathcal{P}^{\mm}_s(\nu_i)=\sum\nolimits_{i=1}^{\textit{L}} f_i \, \p[\gamma_{m_{x_i}} \geq \underbrace{2^{N_{x_i}\nu_i} - 1}_{Q_i}]\\
&=\sum\nolimits_{i=1}^{\textit{L}} f_i \,\p\Bigg[\dfrac{\mathrm{P}_m G_{x_i} \mathcal{X}_{x_i} r^{-\alpha_j}_{x_i}}{ \Bigg(\sigma^2_{m_{x_i}} + \underbrace{\sum_{t\in \Phi^c_{m_i}} \mathrm{P}_m G_t \mathcal{X}_t r^{-\alpha_{\hat{j}}}_t}_{I^{\hat{G},\hat{j}}_{\Phi^c_{m_i}}}\Bigg)} \geq Q_i\Bigg]\\
&\overset{(a)}{\leq} \sum\nolimits_{i=1}^{\textit{L}} f_i \, \E_{\substack{ \alpha_j, \,I^{\hat{G},\hat{j}}_{\Phi^c_{m_i}}}} \Bigg\{ 1 - \Bigg[ 1- \textrm{exp} \Bigg( \dfrac{-AQ_i(\sigma^2_{m_{x_i}}+I^{\hat{G},\hat{j}}_{\Phi^c_{m_i}})}{\mathrm{P}_m G_{x_i} r^{-\alpha_j}_{x_i}} \Bigg)\Bigg]^{\hat{m}}\Bigg\}\\
&\overset{(b)}{=} \sum\nolimits_{i=1}^{\textit{L}} f_i \, \E_{\substack{ \alpha_j,\,\,I^{\hat{G},\hat{j}}_{\Phi^c_{m_i}}}} \Bigg\{1- \sum\nolimits_{l=0}^{\hat{m}} \binom{\hat{m}}{l}(-1)^l \textrm{exp} \Bigg( \dfrac{-AlQ_i(\sigma^2_{m_{x_i}}+I^{\hat{G},\hat{j}}_{\Phi^c_{m_i})}}{\mathrm{P}_m G_{x_i} r^{-\alpha_j}_{x_i}}\Bigg)\Bigg\}\nonumber\\
&\overset{(c)}{=} \sum\nolimits_{i=1}^{\textit{L}}f_i \sum\nolimits_{l=1}^{\hat{m}}\binom{\hat{m}}{l}(-1)^{l+1}\E_{\substack{I^{\hat{G},\hat{j}}_{\Phi^c_{m_i}}}} \left \{ \sum\nolimits_{j\in\{\LOS,\NLOS\}}  p_j  \textrm{exp} \Bigg( \dfrac{-AlQ_i(\sigma^2_{m_{x_i}}+I^{\hat{G},\hat{j}}_{\Phi^c_{m_i})}}{\mathrm{P}_m G_{x_i} r^{-\alpha_j}_{x_i}}\Bigg) \right \}\\
&\overset{(d)}{=}\sum\nolimits_{i=1}^{\textit{L}}f_i \sum\nolimits_{l=1}^{\hat{m}}\binom{\hat{m}}{l}(-1)^{l+1} \sum\nolimits_{j\in\{\LOS,\NLOS\}}  p_j   \E_{\substack{I^{\hat{G},\hat{j}}_{\Phi^c_{m_i}}}}\left\{\textrm{exp} \Bigg( \dfrac{-AlQ_i(\sigma^2_{m_{x_i}}+I^{\hat{G},\hat{j}}_{\Phi^c_{m_i})}}{\mathrm{P}_m G_{x_i} r^{-\alpha_j}_{x_i}}\Bigg) \right \}\\
&=\sum_{i=1}^{\textit{L}}f_i \sum_{l=1}^{\hat{m}}\binom{\hat{m}}{l}(-1)^{l+1} \sum_{j\in\{\LOS,\NLOS\}}  p_j  \textrm{exp}\Bigg( \dfrac{-AlQ_i \sigma^2_{m_{x_i}}}{\mathrm{P}_m G_{x_i} r^{-\alpha_j}_{x_i}}\Bigg) \E_{\substack{I^{\hat{G},\hat{j}}_{\Phi^c_{m_i}}}}\left\{\textrm{exp} \Bigg( \dfrac{-AlQ_i I^{\hat{G},\hat{j}}_{\Phi^c_{m_i}}}{\mathrm{P}_m G_{x_i} r^{-\alpha_j}_{x_i}}\Bigg) \right \}\,. \label{eq:33}
\end{align}
Here, $x_i$ refers to the associated mmWave BS that has the requested file $i$ in its local cache. 
Particularly, (a) follows from the tight {lower} bound of a Gamma random variable \cite{MyListOfPapers:Thornburg_TSP_2014}, and by taking unconditional expectation with respect to path loss exponent $\alpha_j$ and interference $I^{\hat{G},\hat{j}}_{\Phi^{c}_{m_i}}$. Further, (b) follows from the Binomial theorem, (c) is  obtained by taking average over $\alpha_j$, and (d) is obtained {by} exploiting the independence of $P_j$ over $I^{\hat{G},\hat{j}}_{\Phi^c_{m_i}}$. Now, 
applying the thinning theorem of a PPP by considering blockages and effective antenna gains, the point process $I^{\hat{G},\hat{j}}_{\Phi^c_{m_i}}$ can be divided into  6 independent sub-PPPs as shown in the following.

\begin{multline}\label{eq: possibilities of antenna gains2}
I^{\hat{G},\hat{j}}_{\Phi^c_{m_i}} = I^{G^{\rm{M}}G^{\rm{M}},\LOS}_{\Phi^c_{m_i}}+I^{G^{\rm{M}}G^{\rm{m}},\LOS}_{\Phi^c_{m_i}}+I^{G^{\rm{m}}G^{\rm{m}},\LOS}_{\Phi^c_{m_i}}+I^{G^{\rm{M}}G^{\rm{M}},\NLOS}_{\Phi^c_{mi}}+I^{G^{\rm{M}}G^{\rm{m}},\NLOS}_{\Phi^c_{m_i}}+I^{G^{\rm{m}}G^{\rm{m}},\NLOS}_{\Phi^c_{m_i}}
\end{multline}
Accordingly, the expectation part in \eqref{eq:33} can be reduced to
\begin{align}
&\mathbb{E}_{I^{\hat{G}_{\hat{q}},\hat{j}}_{\Phi^c_{m_i}}}\left\{\textrm{exp}\left(\frac{-AlQ_i (I^{G^{\rm{M}}G^{\rm{M}},\LOS}_{\Phi^c_{m_i}}+I^{G^{\rm{M}}G^{\rm{m}},\LOS}_{\Phi^c_{m_i}}+I^{G^{\rm{m}}G^{\rm{m}},\LOS}_{\Phi^c_{m_i}}+I^{G^{\rm{M}}G^{\rm{M}},\NLOS}_{\Phi^c_{mi}}+I^{G^{\rm{M}}G^{\rm{m}},\NLOS}_{\Phi^c_{m_i}}+I^{G^{\rm{m}}G^{\rm{m}},\NLOS}_{\Phi^c_{m_i}})}{\mathrm{P}_m G_{x_i} r^{-\alpha_\LOS}_{x_i}}\right) \right\}\nonumber\\
&{=}\prod_{\hat{q}=1}^{3} \prod_{\hat{j}\in\{\LOS,\NLOS\}}\mathbb{E}_{I^{\hat{G}_{\hat{q}},\hat{j}}_{\Phi^c_{m_i}}}\left\{\textrm{exp}\left(\frac{-AlQ_i I^{\hat{G}_{\hat{q}},\hat{j}}_{\Phi_{m_i}}}{\mathrm{P}_m G_{x_i} r^{-\alpha_j}_{x_i}}\right)\right\}\label{eq27}\,,
    \end{align}
    where \eqref{eq27} follows from the fact that sub-PPPs in \eqref{eq: possibilities of antenna gains2} are independent. 
    Below we compute the expectation of 
$I^{G^{\rm{M}}G^{\rm{M}},\LOS}_{\Phi_{m_i}}$ only. All other terms in \eqref{eq27} can be derived in a similar way. By utilizing Laplace transform, the above expectation can be given as
\begin{align}
&\mathbb{E}\left[\textrm{exp}\left(-s_{(i,l,j)}I^{\hat{G},\hat{j}}_{\Phi^c_{m_i}}\right)\big|\hat{G}=G^{\rm{M}}G^{\rm{M}}, \hat{j}=\LOS\right] \nonumber \\
&\overset{(f)}{=} \mathbb{E}_{\Phi^{\hat{G}_1,\LOS}_{\overline{m}_i}} \Big\{ \prod_{t\in \Phi^{\hat{G}_1,\LOS}_{\overline{m_i}}} \mathbb{E}_{\mathcal{X}_t} \Big[ \textrm{exp}(-s_{(i,l,j)} \mathrm{P}_m G^{\rm{M}}G^{\rm{M}} \mathcal{X}_t r^{-\alpha_\LOS}_t )\Big] \Big\}\nonumber\\
&\overset{(g)}{=}\mathbb{E}_{\Phi^{\hat{G}_1,\LOS}_{\overline{m}_i}} \Big\{ \prod_{t\in \Phi^{\hat{G}_1,\LOS}_{\overline{m_i}} }  \Big( \frac{1}{1+\frac{s_{(i,l,j)} \mathrm{P}_m G^{\rm{M}}G^{\rm{M}} r^{-\alpha_\LOS}_t }{\hat{m}} }\Big)^{\hat{m}} \Big\}\\
&\overset{(h)}{=}\textrm{exp}\Big[-\overline{\lambda}_{m_i}  \int_{0}^{\infty}\Big(1-\Big(\frac{1}{1+\frac{s_{(i,l,j)} \mathrm{P}_m G^{\rm{M}}G^{\rm{M}} r^{-\alpha_\LOS}_t }{\hat{m}}}\Big)^{\hat{m}}\Big) e^{-\beta r_t} 2 \pi  p_{\rm{MM}} r_t d r_t\Big]\,,
\end{align}
where $\Phi^c_{m_i}$ and $\Phi_{\overline{m}_i}$ have the same meaning but the use of the later is for notational simplicity. Further, $\Phi^{\hat{G}_1,\hat{j}}_{\overline{m}_i}$ is the point process marked by {the unavailability of file $i$}, $G^MG^M$ and LoS state. $s_{(i,l,j)} =\frac{AlQ_i}{\mathrm{P}_m G_{x_i} r^{-\alpha_j}_{x_i}} \Big \vert_{\substack{G_{x_i} = G^{\rm{M}}G^{\rm{M}}}} = \frac{AlQ_i}{\mathrm{P}_m (G^{\rm{M}}G^{\rm{M}}) r^{-\alpha_j}_{x_i}}$ and $p_{\rm{MM}}$ is the probability when the antenna gain takes value $G^{\rm{M}}G^{\rm{M}}$ and $e^{-\beta r_k}$ is the probability of the occurrence of LoS transmission.  In the above, (f) follows from independent  channel fading gains, 
(g) follows from the moment generating function of a Nakagami-$\hat{m}$ random variable, and (h) follows from the probability generating functional of a PPP \cite{MyListOfPapers:HaenggiBook2012}.

Now, taking into account all the sub-PPPs given by \eqref{eq: possibilities of antenna gains2}, we have 
\begin{align}
&\E[\textrm{exp}(-s_{(i,l,j)} I^{\hat{G},\hat{j}}_{\Phi^c_{m_i}})]=\E[\textrm{exp}(\sum\nolimits_{\hat{q}=1}^{3}\sum\nolimits_{\hat{j} \in \{L,N\}}(-s_{(i,l,j)} I^{\hat{G}_{\hat{q}},\hat{j}}_{\Phi^c_{m_i}}))]  \\
&\overset{(i)}=\prod_{\hat{q}=1}^{3}\prod_{\hat{j} \in \{\LOS,\NLOS\}}\E[\textrm{exp}(-s_{(i,l,j)} I^{\hat{G}_{\hat{q}},\hat{j}}_{\Phi^c_{m_i}})]\\
&=\prod_{\hat{q}=1}^{3}\prod_{\hat{j} \in \{\LOS,\NLOS\}}\textrm{exp}\Big[-\overline{\lambda}_{m_i} \int_{0}^{\infty}\Big(1-\Big(\frac{1}{1+\frac{s_{(i,l,j)} \mathrm{P}_m \hat{G}_{\hat{q}} r^{-\alpha_{\hat{j}}}_t }{\hat{m}}}\Big)^{\hat{m}}\Big) p_{\hat{j}} 2 \pi p_{\hat{q}} r_t d r_t\Big]\,,
\end{align}
where (i)  follows from the fact that all sub-PPPs are independent.
 Finally, substituting the above results into \eqref{eq:33}, the desired proof is obtained.}
\vspace{-1.25em}
\section{Proof of Theorem 4}\vspace{-.75em}
Considering the NL scenario, the received  SNR at the typical user in the mmWave network served by BS $x_i$ can be reduced to a form given as\vspace{-0.5em}
	\begin{align}
        \gamma^{\textrm{SNR}}_{m_{x_i}} \approx \left({\mathrm{P}_m \, G_{x_i} \, \mathcal{X}_{x_i} \, r^{-\alpha_j}_{x_i}}\right)/\left({\sigma^2_{m_{x_i}}}\right). \label{eq2}
    \end{align}
	Accordingly, the rate supported by the serving BS delivering file $i$ to the typical user is\vspace{-0.5em}
	\begin{align}
        R^{\textrm{SNR}}_{m_{x_i}}\approx  \textrm{log} (1 + \gamma^{\textrm{SNR}}_{m_{x_i}})/N_{x_i}. \label{eq3}
	\end{align}
	Now, the conditional probability of file delivery in the mmWave network is given by\vspace{-0.50em}
	\begin{align}
	    \mathcal{P}^{mm}_s (\{\nu_i\}) =&\sum\nolimits_{i=1}^{L} f_i \mathbb{P}[\textrm{log}(1+\gamma^{\textrm{SNR}}_{m_{x_i}})\geq {\underbrace{N_{x_i} \nu_i}_{\rho_i}}] \nonumber \\
	    =& \sum\nolimits_{i=1}^{L} f_i \mathbb{P}[\frac{\mathrm{P}_m G_{x_i} \mathcal{X}_{x_i} r^{-\alpha_j}_{x_i}}{\sigma^2_{m_{x_i}}} \geq \underbrace{2^{\rho_i} - 1}_{Q_i}]
        =\sum\nolimits_{i=1}^{L} f_i \mathbb{P}[\frac{r^{\alpha_j}_{x_i}}{\mathcal{X}_{x_i}} \leq  \frac{\mathrm{P}_m G_{x_i}}{Q_i \sigma^2_{m_{x_i}}}],
	    \label{eq4}
	\end{align}
	where $\alpha_j=\{\alpha_{\LOS},\alpha_{\NLOS}\}$.
According to thinning theorem, the PPP $\Phi_{m_i}$ is thinned from the process $\Phi_m$ with density $\lambda_m p_{m_i}$, where $p_{m_i}$ is the probability that the $i$th file is stored in the cache. 
Further, the process $\Phi_m$ consists of two sub-PPPs (\emph{i.e.,} $\Phi^\LOS_{m_i}$ and $\Phi^\NLOS_{m_i}$).
Initially, it is necessary to calculate the density of $\Psi_i= \{r^{\alpha_j}_{x_i}(\triangleq ||\psi_q||)\} = \{||y_{i,q}||^{\alpha_j}\}$, where $i \in \mathcal{S}, q \in \mathbb{N}, j \in \{\LOS,\NLOS\}$. By using the mapping theorem, the intensity measure of the process $\Psi_i$ is given by\vspace{-0.5em}
    \begin{align}
&{\Psi_i}([0,\psi] )= \sum\nolimits_{j\in \{\LOS,\NLOS\}}\int_{0}^{\psi^{\frac{1}{\alpha_j}}} \lambda_m p_j p_{m_i} 2 \pi r \textrm{d}r\,, \\ 
        &=\sum\nolimits_{j\in\{L,N\}} c_j 2 \pi \lambda_m p_{m_i} (-\frac{1}{\beta})\bigg\{\psi^{\frac{1}{\alpha_j}}\textrm{exp}(-\beta \psi^{\frac{1}{\alpha_j}})+\frac{1}{\beta}(\textrm{exp}\big(-\beta \psi^{\frac{1}{\alpha_j}})-1\big)\bigg\}+\lambda_m p_{m_i} \pi \psi^{\frac{2}{\alpha_\NLOS}}\,,\nonumber 
    \end{align}
    where $c_j \in \{-1,1\}, c_\LOS =1, c_\NLOS=-1$. Then the density is given by
    \begin{align}
        &\lambda_{\Psi_i}(\psi)=\frac{\textrm{d} {\Psi_i}([0,\psi]) }{\textrm{d} \psi} \nonumber \\
        &\overset{(a)}=\sum_{j\in\{\LOS,\NLOS\}} \left\{c_j \lambda_m p_{m_i} 2 \pi (-\frac{1}{\beta}) \Big\{(\frac{1}{\alpha_j}) \psi^{(\frac{1}{\alpha_j}-1)} \textrm{exp}(-\beta \psi^{\frac{1}{\alpha_j}}) + \psi^{\frac{1}{\alpha_j}} \textrm{exp}(- \beta \psi^{\frac{1}{\alpha_j}}) (-\beta \frac{1}{\alpha_j}) \psi^{(\frac{1}{\alpha_j}-1)} \Big\} \right.\nonumber \\  &\left.+ \lambda_m p_{m_i} 2 \pi (-\frac{1}{\beta^2} ) \textrm{exp}(-\beta \psi^{\frac{1}{\alpha_j}}) \big(-\beta \frac{1}{\alpha_j} \psi^{(\frac{1}{\alpha_j} - 1)}\big) \right\}+\lambda_m p_{m_i} \pi \frac{2}{\alpha_\NLOS} \psi^{\frac{2}{\alpha_\NLOS}-1} \nonumber
            \end{align}
            \begin{align}
        &=\sum\nolimits_{j\in\{\LOS,\NLOS\}}c_j\lambda_m p_{m_i} 2 \pi (\frac{1}{\alpha_j}) \textrm{exp}(-\beta \psi^{\frac{1}{\alpha_j}})\psi^{(\frac{2}{\alpha_j}-1)}+\lambda_m p_{m_i} \pi \frac{2}{\alpha_\NLOS} \psi^{\frac{2}{\alpha_\NLOS}-1}\nonumber \\
        &\overset{(b)}=\sum\nolimits_{j\in\{\LOS,\NLOS\}}c_j\lambda_m p_{m_i}  \pi \delta_{m_j} \textrm{exp}(-\beta \psi^{\frac{\delta_{m_j}}{2}})\psi^{(\delta_{m_j}-1)}+\lambda_m p_{m_i} \pi \delta_{m_\NLOS} \psi^{\delta_{m_\NLOS}-1}.
    \end{align}
Here, (a) follows from substituting $p_\LOS = \textrm{exp}(-\beta r_{x_i}), p_\NLOS = 1 - \textrm{exp}(-\beta r_{x_i})$ and (b) follows from the fact that $\delta_{m_j} = \frac{2}{\alpha_j}$.
Now, the density of the process $\Omega_i = \{\frac{r^{\alpha_j}}{\mathcal{X}_{x_i}} (\triangleq \frac{||y_{i,q}||^{\alpha_j}}{\mathcal{X}_{x_i}})\}=\{||\omega_q||\}$ is acquired according to the displacement theorem. Below we show its derivation. But, first we give the joint probability of $\psi_q$ and $\mathcal{X}_{x_i}$ as
    \begin{align}
    \mathbb{P}[\frac{\psi}{\mathcal{X}_{x_i}} \leq \omega] &= \mathbb{P}[\mathcal{X}_{x_i}\geq \frac{\psi}{\omega} ]=1-F_{\mathcal{X}}(\frac{\psi}{\omega}).
    \end{align}
Due to the fact that the integral of pdf is its cdf, the joint probability is given by
    \begin{align}
        f(\psi,\omega ) =& \frac{\mathrm{d}\big(1-F_{\mathcal{X}}(\frac{\psi}{\omega})\big)}{\mathrm{d} \omega }=\frac{\psi}{\omega^2}f_{\mathcal{X}}(\frac{\psi}{\omega})=\frac{\psi}{\omega^2} \frac{{\hat{m}}^{\hat{m}} (\frac{\psi}{\omega})^{\hat{m}-1} \textrm{exp}(-\hat{m} (\frac{\psi}{\omega}))}{\Gamma(\hat{m})}.
    \end{align}
According to the displacement theorem, we use the above joint probability to calculate the density of the process $\Omega_i$, which is given by
    \begin{align}
        \lambda_{\Omega_i}(\omega)& = \int_{0}^{\infty} \lambda(\psi) f(\psi,\omega) \textrm{d}\psi\nonumber \\
        &=\int_{0}^{\infty}\Bigg(\sum\nolimits_{j\in\{\LOS,\NLOS\}} c_j \pi \lambda_m \textrm{exp}(-\beta \psi^{\frac{\delta_{m_j}}{2}}) p_{m_i} \delta_{m_j} \psi^{(\delta_{m_j}-1)} + \lambda_m p_{m_i} \pi \delta_{m_\NLOS} \psi^{\delta_{m_\NLOS}-1}\Bigg)\nonumber \\ &\times\frac{\psi {\hat{m}}^{\hat{m}} (\frac{\psi}{\omega})^{\hat{m}-1} \textrm{exp}\big(-\hat{m} (\frac{\psi}{\omega})\big)}{\omega^2 \Gamma(\hat{m})}\textrm{d}\psi\nonumber \\
                &=\sum\nolimits_{j\in\{\LOS,\NLOS\}} c_j \frac{p_{m_i} \lambda_m  \delta_{m_j} {\hat{m}}^{\hat{m}} \pi}{\omega^{(\hat{m}+1)}\Gamma(\hat{m})} \int_{0}^{\infty} \psi^{(\delta_{m_j}+\hat{m}-1)}\textrm{exp}(-\frac{\hat{m}}{\omega} \psi -\beta \psi^{\frac{\delta_{m_j}}{2}})\textrm{d}\psi \nonumber\\
        &+ p_{m_i} \lambda_m \pi \delta_{m_\NLOS}\omega^{\delta_{m_\NLOS}-1}\frac{\Gamma(\delta_{m_\NLOS} +\hat{m})}{{\hat{m}}^{\delta_{m_\NLOS}}\Gamma(\hat{m})}.\label{10}
    \end{align}
Now, according to the complementary void function, the CDF of $\omega_q$ can be given as
    \begin{align}
       F_{\Omega_i}(\tilde{\omega}) = \mathbb{P}[\omega_q < \tilde{\omega}] = 1 - \mathbb{P}\big[\Omega_i[0,\tilde{\omega})=0\big].
    \end{align}
Since the displacement theorem and mapping theorem of a PPP is still a PPP, $\mathbb{P}[\Omega_i[0,\tilde{\omega})=0]=\textrm{exp}[-\int_{0}^{\tilde{\omega}} \lambda_{\Omega_i}(\omega) \textrm{d}\omega]$. Accordingly, the CDF of $\omega_q$ can be given as
    \begin{align}
        &F_{\Omega_i}(\tilde{\omega})
        =1-\textrm{exp}\Big(-\sum\nolimits_{j\in\{\LOS,\NLOS\}}k_j p_{m_i} Z_j(\tilde{\omega})-\hat{k}p_{m_i}\tilde{\omega}^{\delta_{m_\NLOS}}\Big),\label{eq11}
    \end{align}
    where $Z_j(\tilde{\omega})= \int_{0}^{\infty}\int_{0}^{\tilde{\omega}}\frac{\textrm{exp}(-\frac{\hat{m}}{\omega}\psi)}{\omega^{(\hat{m}+1)}}\textrm{d}\omega \psi^{(\delta_{m_j}+\hat{m}-1)} \textrm{exp}(-\beta \psi^{\frac{\delta_{m_j}}{2}})\textrm{d}\psi$ and should be a positive value due to the form of the integrand, $\hat{k} = \pi \lambda_m \frac{\Gamma(\delta_{m_\NLOS}+\hat{m})}{\Gamma(\hat{m}){\hat{m}}^{\delta_{m_\NLOS}}}$, and $k_j = c_j\frac{\pi \lambda_m {\hat{m}}^{\hat{m}}\delta_{m_j}}{\Gamma(\hat{m})}$.
    Now, according to \eqref{eq4} and \eqref{eq11}, the ASP of file delivery in mmWave network can be written as    
    \begin{align}
         \mathbb{P}[\gamma^{\textrm{SNR}}_{m_{x_i}}\geq Q_i]\!= \!\!F_{\Omega_i}(\frac{\mathrm{P}_m G_{x_i}}{Q_i \sigma^2_{m_{x_i}}}) \!=\! 1 \!-\! \textrm{exp}\bigg(\!\!\!-\!\!\sum_{j\in\{\LOS,\NLOS\}} k_j p_{m_i} Z_j\big(\frac{\mathrm{P}_m G_{x_i}}{Q_i \sigma^2_{m_{x_i}}}\big)-\hat{k} p_{m_j} (\frac{\mathrm{P}_m G_{x_i}}{Q_i \sigma^2_{m_{x_i}}})^{\delta_{m_\NLOS}}\bigg)
    \end{align}
Finally, we can generate the overall probability from \eqref{eq4} as
    \begin{align}
        \mathcal{P}^{mm}_s(\nu_i)
        &=\sum\nolimits_{i=1}^{L}f_i\Big[1 - \textrm{exp}\bigg(-\sum\nolimits_{j\in\{\LOS,\NLOS\}}k_j p_{m_i} Z_j\big(\frac{\eta_i G_{x_i}}{Q_i }\big)-\hat{k}p_{m_i}\big(\frac{\eta_i G_{x_i}}{Q_i}\big)^{\delta_{m_\NLOS}}\bigg)\Big]\,,
    \end{align}
    where $k_j=c_j\frac{\pi  \lambda_m {\hat{m}}^{\hat{m}} \delta_{m_j}}{\Gamma(\hat{m})}, \eta_i = \frac{\mathrm{P}_m}{\sigma^2_{m_{x_i}}}, p_j\in\{p_\LOS, p_\NLOS\}, \delta_{m_j} = \frac{2}{\alpha_j}, \forall j \in \{\LOS,\NLOS\}$.
   
\vspace{-.5em}
\bibliographystyle{IEEEtran}\vspace{-.75em}
\bibliography{IEEEabrv,MyListOfPapers}
\end{document}